\documentclass[fleqn,usenatbib]{mnras}
\usepackage{newtxtext,newtxmath}
\usepackage[T1]{fontenc}

\DeclareRobustCommand{\VAN}[3]{#2}
\let\VANthebibliography\thebibliography
\def\thebibliography{\DeclareRobustCommand{\VAN}[3]{##3}\VANthebibliography}

\usepackage{threeparttable}
\usepackage{physics}
\usepackage{graphicx}
\usepackage{amsmath}
\usepackage{comment}
\usepackage{xcolor}
\usepackage[normalem]{ulem}

\usepackage{tikz-cd}
\usepackage{multicol}
\usepackage{multirow}
\usepackage{array}
\usetikzlibrary{arrows.meta,decorations.markings}

\usepackage[abs]{overpic}

\newcommand{\Sun}{_{\odot}}
\newcommand{\valrango}[2]{\begin{tabular}[c]{@{}c@{}}#1\\\footnotesize#2\end{tabular}}

\title[Tidal disruption of stellar binaries]{Tidal disruption of stellar binaries as a pathway to exotic transients}
\author[M.~González-Servín et al.]{ Mauricio
González-Servín,$^{1}$\thanks{E-mail: m.gonzalez@irya.unam.mx}
Emilio Tejeda$^{2}$, Susana Lizano$^{1}$, Luis A.~Manzaneda$^{1}$
\newauthor
and Raúl Cano-Villegas$^{3}$
\\
$^{1}$Instituto de Radioastronom\'ia y Astrof\'isica, Universidad Nacional Aut\'onoma de M\'exico, Antigua Carretera a P\'atzcuaro $\#$ 8701, Ex-Hda. San Jos\'e de la \\ Huerta, Morelia, Michoac\'an, C.P. 58089, M\'exico\\
$^{2}$SECIHTI--Instituto de F\'{i}sica y Matem\'{a}ticas, Universidad Michoacana de San Nicol\'{a}s de Hidalgo, Ciudad Universitaria, 58040 Morelia, Mich., M\'exico\\
$^{3}$Facultad de Ciencias Físico Matem\'{a}ticas, Universidad Michoacana de San Nicol\'{a}s de Hidalgo, Ciudad Universitaria, 58040 Morelia, Mich., M\'exico
}

\date{Accepted XXX. Received YYY; in original form ZZZ}

\pubyear{\the\year{}}

\begin{document}
\label{firstpage}
\pagerange{\pageref{firstpage}--\pageref{lastpage}}
\maketitle

\begin{abstract}

Tidal disruption event (TDE) progenitors are commonly modelled as single stars on parabolic orbits around a supermassive black hole (SMBH), yet observations reveal a richer diversity of dynamical pathways. We show that tidal separation of stellar binaries by a $10^6\,M_\odot$ SMBH provides a natural mechanism for producing eccentric TDEs. Using restricted three-body and SPH simulations, we model a solar-like star (SLS)--white dwarf (WD) binary on a parabolic orbit. The binary orbital phase governs the outcome: one component is captured onto a tightly bound orbit while the other is ejected as a hypervelocity object, naturally producing TDEs with eccentricities $e \neq 1$. We classify these into Elliptical TDEs (eTDEs) and Hyperbolic TDEs (hTDEs), occurring with equal probability. For $\sim 88\%$ of orientations the disruption is clean, with no mass accreted by the WD. The remaining $\sim 12\%$, in two narrow phase windows, leaves the WD with a captured debris envelope (WDDE). About a third of that, $\sim 4\%$ of orientations, involves a direct WD--SLS collision near pericenter, with fallback peaking up to five times earlier and twenty times higher than for a single star; for the innermost $\sim 1.5\%$ the total WDDE mass exceeds $1.4\,M_\odot$, though the degenerate core does not, since the captured material forms a non-degenerate envelope, suggesting outcomes ranging from nova-like events to peculiar red giant-like objects. Depending on the phase, the mechanism may also produce repeating partial TDEs (rTDEs) and quasi-periodic eruptions (QPEs). Binary--SMBH encounters provide a robust channel for generating diverse TDEs with distinct observational signatures.
\end{abstract}



\section{Introduction}
\label{sec:intro}

Tidal disruption events (TDEs) occur when a star ventures too close to a supermassive black hole (SMBH) and is ripped apart by its tidal forces. In the classical TDE picture, a star on a parabolic orbit is disrupted, leading to a fallback rate of stellar debris that scales at late times as \( t^{-5/3} \) and powers a luminous flare \citep{rees1988tidal, evans1989tidal}. This standard model has successfully explained numerous observed TDE candidates detected across various wavelengths, from soft X-rays to the ultraviolet and optical regimes \citep{bade1996detection, esquej2007candidate, gezari2012ultraviolet,Vanvelzen2021ApJ,langis2026repeating}.

Recently, theoretical studies have explored eccentric TDEs, where the disrupted star is on a bound elliptical orbit rather than a parabolic one. These encounters significantly alter the characteristic fallback rate curve, leading to higher peak accretion rates, steeper late-time slopes, and for sufficiently bound orbits, a finite-duration flare rather than a power-law decay \citep{Cufari_2022, Liu_2023, Hu_2024}. The distinct observational signatures of eccentric TDEs make them particularly interesting, as they could explain unusual transient phenomena that deviate from classical TDE predictions. However, a fundamental challenge remains: identifying a natural astrophysical mechanism that can place a single star on such a disruptive, tightly-bound orbit around an SMBH.

Binary star systems, which represent the norm rather than the exception in the field \citep{Tian2018},\footnote{The binary fraction is expected to be appreciably lower in a nuclear star cluster than in the field, since frequent close encounters separate the softer pairs \citep[e.g.,][]{alexander2005stellar,2016MRASALEXANDER,stone2020rates}. The system studied here, with $a_0 = 11\,R_\odot$, is hard against separation in such an environment. We return to this suppression factor in Section~\ref{subsec:rates}.} offer a promising natural channel for these exotic events. As a binary interacts with an SMBH, the three-body dynamics can lead to tidal separation of the pair, potentially ejecting one star as a hypervelocity object while capturing the other onto a tightly-bound, elliptical orbit \citep{hills1988hyper,brown2015hypervelocity}. This scenario provides a natural pathway to eccentric TDEs, while also enabling a rich variety of exotic outcomes \citep{mandel2015double,yu2024binary}. Initial studies of binary-SMBH encounters have mainly treated stars as point particles, overlooking the crucial hydrodynamic responses that occur during close encounters \citep{kobayashi2009hypervelocity, addison2019busting}.

Previous hydrodynamical simulations have begun to explore these outcomes. \citet{antonini2011tidal} extended point-mass studies by treating stars as fluids, finding that encounters can result in stellar collisions, hypervelocity ejections, and long-term capture of stars in tight orbits around the SMBH. \citet{mainetti2016hydrodynamical} investigated tidal stripping of binaries using smoothed particle hydrodynamics (SPH) simulations, revealing that accretion luminosities often display a characteristic ``knee'' rather than double-peaked light curves, particularly in unequal-mass systems. In the context of multiple tidal events, \citet{bonnerot2019streams} showed that in a double TDE the two debris streams may collide before falling back to the SMBH, producing a short optical precursor to the main flare; when the double disruption follows the tidal separation of a binary, they find collision probabilities as high as $44\%$. More recently, \citet{ryu2023close} focused on encounters between tight binaries and stellar-mass BHs that lead to disruptions, runaway stars, or X-ray binary formation, while \citet{yu2025binary} simulated binaries of two solar-type stars, finding that collisions during the encounter produce either mergers or highly perturbed remnants depending on encounter depth and geometry, the latter being vulnerable to future partial disruptions and capable of producing signals that resemble classical TDEs but with distinct temporal and morphological properties. However, a systematic investigation of how binary encounters with SMBHs can specifically produce eccentric TDEs remains lacking.

In this work, we systematically investigate whether binary tidal separation can naturally produce eccentric TDEs and other exotic transients. Using a combination of restricted three-body dynamics and hydrodynamical simulations with the Phantom code \citep{price2018phantom}, we model encounters between a \(10^6 M_\odot\) SMBH and a binary system composed of solar-like star (SLS) and a white dwarf (WD). Our approach reveals that the binary phase is a critical parameter governing the encounter outcome. We identify a specific range of phases that preferentially produces highly eccentric TDEs, along with other remarkable phenomena including stellar collisions, partial TDEs where a stellar core bound to the SMBH survives, and significant mass accumulation onto the WD companion. Most notably, we find scenarios where the combined mass of the WD and its acquired envelope exceeds $1.4\,M_\odot$, although the degenerate core itself does not, pointing to a range of possible outcomes including nova-like episodes or peculiar red giant-like configurations. The diversity of outcomes uncovered in our simulations demonstrates that binary encounters with SMBHs represent a fertile pathway for generating exotic transients with distinct observational signatures.

This paper is organized as follows. In Section~\ref{sec:theory} we present the general theoretical framework that motivates this work. We first review classical TDEs and their extension to eccentric events, and then discuss the separation of stellar binary systems due to the tidal forces exerted by a SMBH.
In Section~\ref{sec:methodology} we describe the methodology followed in this work and the configuration of the parameters used in our simulations. In this section we also present the initial conditions of the problem, the relativistic considerations included in the model, and a preliminary study of the three-body problem.
In Section~\ref{sec:dynamics} we analyse the general behaviour of the hydrodynamical simulations, including both their dynamical evolution and their overall morphology. In Section~\ref{sec:WDDE} we focus on studying the possible final fate of the WD.
In Section~\ref{sec:menagerie} we investigate the different types of TDEs that can be produced in our scenario, including elliptical, hyperbolic, and collision driven events.
In Section~\ref{sec:summary} we discuss the implications of our results and their possible connection with observational signatures.
Finally, in Section~\ref{sec:conclusions} we summarize the main conclusions of this work.

\section{Theoretical framework}
\label{sec:theory}

\subsection{Classical tidal disruptions}
\label{sec:theory_classical}
One of the most important assumptions in the classical TDE scenario is that the star approaching the black hole (BH) follows a parabolic orbit. This represents an excellent approximation for individual stars scattered into loss-cone orbits from outside the BH's sphere of influence through multi-body gravitational interactions \citep{Merritt2010}.

The tidal radius for a single star is defined as the distance from the BH at which the tidal forces overcome the star's self-gravity. For a star of mass $m_s$ and radius $r_s$ approaching a BH of mass $M_{\rm BH}$, this is given by:
\begin{equation}
\label{eq:tidal_radius_single}
r_t = \left( \frac{M_{\rm BH}}{m_s} \right)^{1/3} r_s.
\end{equation}

Equation~\eqref{eq:tidal_radius_single} is an order-of-magnitude estimate: a form factor set by the internal density profile introduces an order-unity correction, upwards for low-mass stars and downwards by up to $\sim2$ above $\sim1\,M_\odot$ \citep{Stone2013,ryu2020tidalI}. Since $m_s$ and $r_s$ are fixed in all our simulations, this is a constant rescaling common to every run and does not affect the phase-dependent trends studied here.

During a classical TDE, approximately half of the star's material remains gravitationally bound to the BH, eventually forming an accretion disk, while the remainder is ejected \citep{rees1988tidal, evans1989tidal}. The bound material returns to pericenter over a range of timescales, producing a fallback rate that follows at late times a power-law \citep{LINDA2009MNRAS,lodato2009stellar}:
\begin{equation}
\label{eq:fallback_rate}
\dot M \approx \frac{1}{3} \dot M_0 \left( \frac{t}{t_0} \right)^{-5/3},
\end{equation}
where the characteristic timescale $t_0$ corresponds approximately to the orbital period of the most tightly bound debris:
\begin{align}
    t_0 & = \pi\sqrt{2} \left(\frac{r_s^3}{Gm_s}\right)^{1/2} \left(\frac{M_{\rm BH}}{m_s}\right)^{1/2} \nonumber \\
    & \simeq 0.22 \,\left(\frac{r_s}{R\Sun}\right)^{3/2} \left(\frac{m_s}{M_\odot}\right)^{-1} \left(\frac{M_{\rm BH}}{10^6 M_\odot}\right)^{1/2} \text{ yr}
\end{align}
and the characteristic accretion rate $\dot{M}_0$ is given by:
\begin{equation}
\dot{M}_{0} = \frac{m_s}{t_0} \simeq 4.46\,\left(\frac{r_s}{R\Sun}\right)^{-3/2} \left(\frac{m_s}{M_\odot}\right)^{2} \left(\frac{M_{\rm BH}}{10^6 M_\odot}\right)^{-1/2} M_\odot\,\text{yr}^{-1}.
\end{equation}

Our primary diagnostic throughout is the mass fallback rate $\dot{M}$, the rate at which bound debris returns to pericenter. We stress that this is not the accretion rate $\dot{M}_{\rm acc}$ onto the BH, and that a complete theory relating the two is still lacking: the mapping depends on the efficiency of circularisation and on mass lost to outflows. Recent global simulations find that only a small fraction of the returning debris is promptly captured \citep{krolik2024follow}, while observed TDE light curves show that the optical/UV luminosity does not simply follow $\dot{M}$ \citep{mummery2025calorimetry}. The luminosities quoted below are obtained by scaling $\dot{M}$ directly (Section~\ref{subsec:cTDE}), and should therefore be read as order-of-magnitude indicators of the energy available rather than as predictions of an observed light curve.

The tidal strength for a tidal encounter is characterized by the impact parameter:
\begin{equation}
\label{eq:impact_parameter_single}
\beta_{s}=\frac{r_{t}}{r_{p}},
\end{equation}
where $r_{p}$ is the pericenter distance between the star and the BH.

Note that $t_0$ and $\dot{M}_0$ above carry no dependence on $\beta_s$. This is the standard ``frozen-in'' result: the spread in specific energy is imprinted when the star crosses the tidal radius rather than at pericenter, so $\beta_s$ cancels \citep{lodato2009stellar,Stone2013}. Simulations nevertheless reveal a residual, weak dependence on $\beta_s$ that this scaling does not capture \citep{guillochon2013hydrodynamical}, of no consequence here since $\beta_s = 1.2$ is held fixed throughout our suite.

We close this section by quantifying the sense in which the parabolic approximation is accurate. The disruption imprints on the debris a spread in specific orbital energy, set by the tidal field across the stellar radius at the moment the star crosses the tidal radius,
\begin{align}
\label{eq:energy_spread}
\Delta E & = \frac{G M_{\rm BH}\, r_s}{r_t^{2}} \nonumber \\
& \simeq 1.9\times10^{17}
\left(\frac{r_s}{R\Sun}\right)^{-1}
\left(\frac{m_s}{M\Sun}\right)^{2/3}
\left(\frac{M_{\rm BH}}{10^{6} M\Sun}\right)^{1/3}
\,\mathrm{cm^{2}\,s^{-2}},
\end{align}
which sets the energy scale of the problem and is used throughout what follows. The relevant reservoir of stars is that contained within the sphere of influence of the SMBH, whose radius is
\begin{equation}
\label{eq:influence_radius}
r_h = \frac{G M_{\rm BH}}{\sigma^{2}} \simeq 0.05
\left(\frac{M_{\rm BH}}{10^{6} M\Sun}\right)
\left(\frac{\sigma}{300\ \mathrm{km\,s^{-1}}}\right)^{-2}\,\mathrm{pc}.
\end{equation}
A star scattered into the loss cone from this radius arrives with a specific orbital energy of order $E \simeq \sigma^{2}/2$, or equivalently with a semi-major axis comparable to $r_h$ itself. Using equation~\eqref{eq:tidal_radius_single} to eliminate $r_t$, the ratio of the two energies is
\begin{equation}
\label{eq:energy_ratio}
\frac{E}{\Delta E} = \frac{1}{2}\,\frac{r_s}{r_h}
\left(\frac{M_{\rm BH}}{m_s}\right)^{2/3}.
\end{equation}
For our fiducial parameters, $m_s = 1\,M_\odot$, $r_s = 1\,R_\odot$ and $M_{\rm BH} = 10^{6}\,M_\odot$, equation~\eqref{eq:energy_ratio} gives $E/\Delta E \simeq 2\times10^{-3}$ at $\sigma = 300$ km s$^{-1}$, and an order of magnitude less at $\sigma = 100$ km s$^{-1}$. The energy the star brings in is therefore negligible compared with the energy spread imparted by the disruption itself, and $E = 0$ is an excellent approximation. The corresponding statement for the eccentricity is simply $|1-e| = r_p/r_h$, which for $r_p \simeq 100\,R_\odot$ ranges from $5\times10^{-6}$ to $5\times10^{-5}$ across the quoted interval of $\sigma$. Both $E = 0$ and $e = 1$ are thus excellent approximations for single-star disruptions \citep{alexander2017}.

\subsection{Eccentric tidal disruptions}
\label{sec:ecc_TDE}

\begin{figure}
    \centering
    \includegraphics[width=\linewidth]{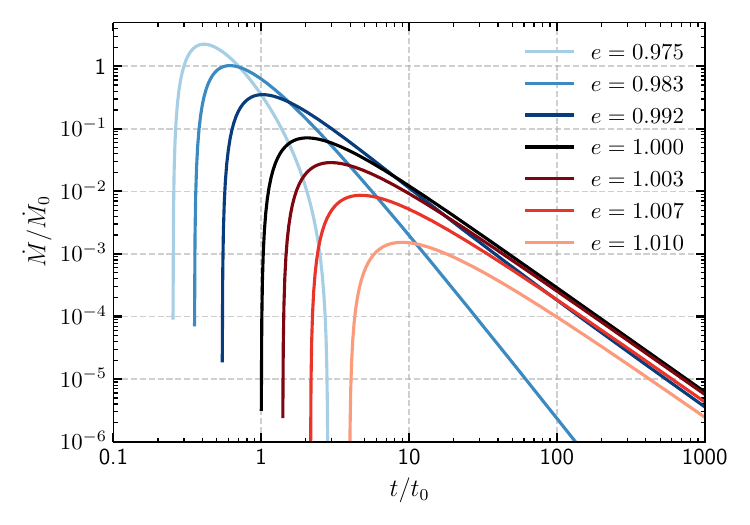}
    \caption{ Mass fallback rates as a function of time for different orbital eccentricities, computed from the illustrative energy distribution of equation~\eqref{eq:energy_distribution}. The remaining model parameters are fixed at $m_s = 1\,M_\odot$, $r_s = 1\,R_\odot$, $M_{\rm BH} = 10^6\,M_\odot$, and $\beta_s = 1.2$, matching the value adopted in all of our simulations. }
    \label{fig:energy_fallback}
\end{figure}

Recently, studies have examined how fallback rate curves change depending on the eccentricity of the disrupted star \citep{Cufari_2022, Liu_2023, Hu_2024}. The shape of the fallback rate curve is determined by the energy distribution of the stellar debris following disruption. As a first-order approximation, we model this using a quartic distribution
\begin{equation}
\label{eq:energy_distribution}
\dv{M}{E} = \frac{15\,m_s}{16\,\Delta E} \left[1 - \left(\frac{E - E_0}{\Delta E} \right)^2\right]^2,
\end{equation}
where $E_0$ is the centre of the distribution and $\Delta E$, defined in equation~\eqref{eq:energy_spread}, represents its width. Equation~\eqref{eq:energy_distribution} is adopted purely as a convenient analytic form with the correct qualitative properties, and not as a quantitative description of the debris energy distribution found in simulations, which is peaked, asymmetric and dependent on the internal structure of the star \citep{lodato2009stellar,guillochon2013hydrodynamical}. It is used only for the illustrative curves of Figure~\ref{fig:energy_fallback}; every fallback rate reported in Sections~\ref{sec:dynamics}--\ref{sec:menagerie} is instead derived from the SPH particle energies themselves.

The width $\Delta E$ can be connected to the orbital eccentricity. The eccentricity of a star moving within the gravitational potential of the BH can be expressed in terms of its specific orbital energy $E_0$ and specific angular momentum $L$ as
\begin{equation}
e^{2} = 1 + \frac{2E_0L^{2}}{(G M_{\rm BH})^{2}}.
\label{eq:eccentricity_single}
\end{equation}
In general, we can express $L$ in terms of $e$ and the pericenter distance $r_p$ as:
\begin{equation}
L = \sqrt{G M_{\rm BH} r_{p} (1+e)},
\label{eq:angular_momentum_single}
\end{equation}
obtaining the expression \citep[e.g.,][]{Hayasaki2018ApJ}
\begin{equation}
e = 1 + \frac{2 E_0}{\beta_s \Delta E } \left(\frac{m_{s}}{M_{\rm BH}}\right)^{1/3}.
\label{eq:eccentricity_final}
\end{equation}

Equations~\eqref{eq:energy_distribution} and \eqref{eq:eccentricity_final} together
determine the fraction of debris that remains bound to the SMBH. Integrating
equation~\eqref{eq:energy_distribution} over $E<0$ gives
\begin{equation}
\label{eq:bound_fraction}
f_{\rm b}(e) = \frac{15}{16}\left(x - \frac{2}{3}x^{3} + \frac{1}{5}x^{5} + \frac{8}{15}\right),
\qquad
x \equiv \frac{\beta_s}{2}\left(\frac{M_{\rm BH}}{m_s}\right)^{1/3}(1-e),
\end{equation}
with $f_{\rm b} = 0$ for $x \leq -1$ and $f_{\rm b} = 1$ for $x \geq 1$. For a parabolic
orbit $x=0$ and $f_{\rm b} = 1/2$, recovering the familiar result that half of the debris
is bound. The transition between complete binding and complete ejection is confined to a
remarkably narrow interval,
\begin{equation}
\label{eq:ecc_window}
|1-e| \leq \frac{2}{\beta_s}\left(\frac{m_s}{M_{\rm BH}}\right)^{1/3} \simeq 0.017
\end{equation}
for our parameters, outside which the debris is either entirely bound or entirely unbound.
The bound fraction is of interest beyond the shape of the fallback curve, since it sets the mass a compact object can accrete in a single disruption and therefore governs the efficiency of BH growth through repeated TDEs \citep{rizzuto2023growth}.
We note that equation~\eqref{eq:bound_fraction} inherits the qualitative character of
equation~\eqref{eq:energy_distribution} and should be read as a guide to the scaling
rather than as a quantitative prediction.

As illustrated in Figure~\ref{fig:energy_fallback}, the resulting fallback rate curves depend critically on the orbital eccentricity. For parabolic encounters ($e=1$), the energy distribution is centred at $E_0=0$, producing the classical $t^{-5/3}$ power-law (Section~\ref{sec:theory_classical}). For elliptic orbits ($e<1$), a larger fraction of the stellar material remains bound to the BH ($E_0<0$), leading to higher peak fallback rates that occur earlier compared to the parabolic case. We refer to these as elliptical TDEs (eTDEs). In the extreme case of low eccentricity encounters where all debris have $E_0<0$, the accretion concludes in a finite time rather than following a power-law decay.

Conversely, hyperbolic TDEs (hTDEs) occur for encounters with $e>1$ and exhibit qualitatively different behaviour: the majority of debris is unbound ($E_0>0$), resulting in subluminous TDEs with lower peak fallback rates that occur later relative to the parabolic baseline.

Correctly identifying these events observationally poses a significant challenge. Except for the extreme elliptical cases that terminate in finite time, the fallback rate of an eTDE or hTDE may be easily misinterpreted as a standard parabolic event involving a different progenitor star mass or a different impact parameter \citep[e.g.,][]{guillochon2009three}. Distinguishing between these scenarios would require either a highly precise measurement of the fallback rate curve shape or independent constraints on the stellar population. Consequently, it is desirable to identify additional observational cues to undoubtedly classify an event as a eTDE or hTDE.

Several mechanisms can place stars on tightly bound orbits around a SMBH. The most widely discussed is tidal capture: a star passing close to the BH without being disrupted excites strong tides at the expense of its orbital energy, which internal viscosity then dissipates, leaving it on a compact, eccentric orbit \citep{fabian1975tidal,press1977formation,cufari2023tidal,rizzuto2023growth}. Because tides are excited efficiently at pericenter distances a few times larger than $r_t$, this channel has a larger cross section than direct parabolic disruption. The Hills mechanism offers a second route, in which the tidal separation of a binary leaves one component bound to the SMBH \citep{hills1988hyper}. In both cases the captured star may later be driven towards its tidal radius, producing an eccentric disruption. The mechanism explored here is a variant of the second, restricted to encounters deep enough that the captured star reaches its own tidal radius on the first passage; this allows the disruption to be followed hydrodynamically, at the cost of a smaller cross section (Section~\ref{subsec:rates}).

\subsection{Binary tidal separation}
\label{sec:binary_TDE}

For a binary system composed of two stars with masses $m_{1}$ and $m_{2}$, initially separated by a distance $a_{0}$, the tidal radius is defined as:
\begin{equation}
\label{eq:tidal_radius_binary}
R_{t}=\left(\frac{M_{\rm BH}}{m_{b}}\right)^{1/3} a_{0},
\end{equation}
where $m_{b}=m_{1}+m_{2}$ is the total mass of the binary system.

The binary impact parameter is defined as:
\begin{equation}
\label{eq:impact_parameter_binary}
\beta_{b}=\frac{R_{t}}{R_{p}},
\end{equation}
where $R_{p}$ is the closest approach distance between the centre of mass (CM) of the binary system and the BH.

For the same reasons that justify the parabolic approximation for individual star TDEs, we adopt a parabolic trajectory for the binary's CM approaching the SMBH. Where we assume that the binary CM velocity dispersion is the same as the typical stellar velocity dispersion in galactic nuclei.

In general, binary tidal separation is expected for encounters with $\beta_{b}\gtrsim 1$. However, it is important to note that $R_{t}$ provides only an order-of-magnitude estimate of where the tidal forces significantly influence the binary dynamics. As discussed previously in the literature \citep{hut1983topology,manzaneda2024relativistic,Aryabrat2026}, a critical factor determining the outcome of such encounters is the spatial orientation (or phase) with which the binary system approaches the central object.

In this work, we aim to explore binary tidal separation as a natural channel to produce eccentric TDEs, where the three-body dynamics during the encounter can inject one star into a tightly-bound elliptical orbit around the SMBH while ejecting its companion.

The tidal binary separation can be preliminarily examined through energy and angular momentum considerations.
Starting from the general relation between orbital energy, angular momentum, and eccentricity introduced in equations~(\ref{eq:eccentricity_single}) and (\ref{eq:angular_momentum_single}), the eccentricities of both the ejected star ($e_{+}$) and the bound star ($e_{-}$) can be written as
\begin{equation}
e_{\pm}^{2}=1+\frac{2E_{\pm}L_{\pm}^{2}}{(GM_{\rm BH})^{2}}.
\label{eq:eccentricity_general}
\end{equation}

Since the CM angular momentum dominates over the internal binary angular momentum, we can approximate:
\begin{equation}
  L_{\pm} \simeq \sqrt{2GM_{\rm BH}\,R_{p}}.
\end{equation}

The energy exchange during the tidal separation can be estimated from the work done by the tidal field over the binary's extent. The average energy scale for the ejected star is:
\begin{equation}
{E_{+}}_{\mathrm{avg}} \simeq \frac{GM_{\rm BH}a_0}{R_t^2},
\end{equation}
representing the typical energy imparted during separation. On the other hand, the maximum attainable energy, reached when the tidal forces act most effectively, scales as:
\begin{equation}
{E_{+}}_{\mathrm{max}} \simeq \frac{GM_{\rm BH}a_0}{R_t^2}\beta_b.
\end{equation}

By energy conservation, and since the binary's CM approaches on a parabolic orbit ($E_{\rm CM} = 0$), the star that remains bound to the BH acquires energies of order $E_{-} \simeq -E_{+}$. Substituting these energy scales into the eccentricity in equation \ref{eq:eccentricity_general} yields the characteristic eccentricity deviations:
\begin{gather}
{e^{2}_{\pm}}_{\mathrm{avg}} \simeq 1 \pm 4\frac{a_0\,R_p}{R_t^2} = 1 \pm 4\left(\frac{m_b}{M_\mathrm{BH}}\right)^{1/3} \beta^{-1}_b,
\label{eq:ecc_avg}\\
{e^{2}_{\pm}}_{\mathrm{max}} \simeq 1 \pm 4\left(\frac{m_b}{M_\mathrm{BH}}\right)^{1/3}.
\label{eq:ecc_max}
\end{gather}
These expressions demonstrate how the binary separation process naturally produces eccentricities that deviate from parabolic orbits.

\section{Methodology and simulation setup}
\label{sec:methodology}

\subsection{Model parameters}
\label{subsec:parameters}

We model a binary star system approaching a SMBH of mass $M_{\rm BH} = 10^{6}M_\odot$ on a parabolic orbit. The binary system consists of a Solar-Like Star (SLS, $m_1 = M\Sun$, $r_1 = R\Sun$) and a White Dwarf (WD, $m_2 = M\Sun$, $r_2 = 0.01\,R\Sun$) in a circular orbit with separation $a_0 = 11 \,R\Sun$. This separation is chosen to maintain dynamical stability while ensuring the tidal deformation of the SLS by the WD is negligible.\footnote{Following the approximation $\delta r_1/r_1 \approx (m_2/m_1)(r_1/a_0)^3$ \citep{sterne1939apsidal,prialnik2009introduction}, $a_0 > 10 \,R_\odot$ yields $\delta r_1/r_1 \lesssim 0.001$.}

This particular pairing is a deliberate simplification. Our aim is to study eccentric TDEs arising from a plausible dynamical channel, and an SLS--WD binary is the simplest configuration that achieves this: with only one extended component there is a single debris stream to follow, whereas an SLS--SLS binary would generically produce two disruptions with potentially overlapping streams, and a WD--WD binary would produce no disruption at all around a SMBH of $10^{6}\,M_\odot$ (Section~\ref{subsec:GR_effects}).

We note that the adopted mass makes the WD atypically massive, the field population peaking near $0.6\,M_\odot$ \citep{kepler2007white}. The sensitivity of our results to this choice is nevertheless limited, since the encounter is governed by dimensionless ratios rather than absolute scales: under a uniform rescaling of all masses by $\lambda$ and all lengths by $\mu$, the ratios $M_{\rm BH}/m_b$, $m_2/m_1$, $a_0/r_1$ and $\beta_s$ are invariant, and only the time and rate normalisations change. Taking $\lambda = 0.6$ and $\mu = 0.66$, the radius of a $0.6\,M_\odot$ main-sequence star, maps our system exactly onto an equal-mass $0.6\,M_\odot$ binary around a $6\times10^{5}\,M_\odot$ SMBH. Relaxing the equal-mass assumption at fixed $M_{\rm BH}$ has a similarly weak effect, changing the controlling parameters at the $10\%$ level. A systematic exploration of the mass ratio is nevertheless left to future work.

According to equations \eqref{eq:tidal_radius_single} and \eqref{eq:tidal_radius_binary}, the characteristic tidal radii for the binary system, the SLS, and the WD are:
\begin{gather}
R_t \simeq 873 \,R_\odot, \label{eq:rt_binary} \\
r_t = 100\,R_\odot, \label{eq:rt_sls} \\
r_{t,{\rm WD}} = R_\odot. \label{eq:rt_wd}
\end{gather}
The last of these assumes $r_2 = 0.01\,R_\odot$, which is generous for a $1\,M_\odot$ WD, so equation~\eqref{eq:rt_wd} is an upper bound.

We fix the SLS impact parameter at $\beta_s = 1.2$, ensuring its disruption while maintaining astrophysical plausibility \citep{alexander2005stellar}. This corresponds to an SLS pericenter distance of $r_p = r_t/\beta_s \simeq 83\,R_\odot$. Taking the approximation $R_p \simeq r_p$, the binary impact parameter is:
\begin{equation}
\label{eq:beta_b_relation}
\beta_b=\beta_s\frac{R_t}{r_t}\simeq 10.5.
\end{equation}
In contrast, the corresponding impact parameter for the WD is negligible ($\beta_{\rm WD} \simeq 0.01$), ensuring it remains well outside its own disruption limit throughout the encounter. This justifies our treatment of the WD as a point mass and enables us to focus the hydrodynamical simulations exclusively on the disruption of the SLS, thereby optimizing the computational resolution for the resulting debris flow.

A summary of all fixed physical and orbital parameters adopted throughout our simulations is provided in Table~\ref{tab:parameters}.

\subsection{General relativistic considerations}
\label{subsec:GR_effects}
While our simulations employ Newtonian gravity, we assess the potential importance of general relativistic (GR) effects given our system parameters. The Schwarzschild radius is
\begin{equation}
R_\mathrm{Sch} = \frac{2GM_{\rm BH}}{c^2} \simeq 4.24\,R\Sun,
\end{equation}
yielding a pericenter-to-horizon ratio of
\begin{equation}
\frac{r_p}{R_\mathrm{Sch}} \simeq 19.6.
\end{equation}

This substantial separation indicates that direct capture by the event horizon is negligible. Note also that $r_{t,{\rm WD}} = 1\,R_\odot < R_\mathrm{Sch}$: a WD deep enough to reach its own tidal radius would cross the horizon and be swallowed whole before being disrupted. Our point-mass treatment is therefore self-consistent for any encounter geometry, and this is also why WD disruptions require an intermediate-mass black hole with $M_{\rm BH} \lesssim 10^{5}\,M_\odot$ rather than a SMBH \citep{maguire2020tidal,SpinAryabrat2026pre}. The system operates in the weak-field regime where leading-order GR corrections are $\mathcal{O}(R_\mathrm{Sch}/r_p) \lesssim 5\%$ \citep{tejeda2017tidal}. While relativistic precession and other post-Newtonian effects may subtly influence the long-term debris evolution and circularization \citep{skadowski2016magnetohydrodynamical,stone2019stellar}, they are not expected to qualitatively alter the bulk disruption dynamics or the primary outcomes discussed in this work. Recent fully relativistic simulations reinforce this, finding the imprint of general relativity on the debris dynamics to be unexpectedly weak even for encounters far more relativistic than ours \citep{chan2026weak}. Our Newtonian treatment therefore provides an appropriate framework for capturing the leading-order hydrodynamical effects during the tidal separation and initial debris evolution.

\subsection{Initial conditions}
\label{subsec:initial_conditions}

The binary CM is initialized at a distance of $R_0 = 30\,R_t$ from the SMBH, a value that ensures numerical convergence of the encounter dynamics \citep{manzaneda2024relativistic}. We orient the parabolic trajectory such that the pericenter lies on the negative $x$-axis at $(-R_p, 0)$. At $R_0$, the initial position and velocity of the CM are:
\begin{align}
&\mathbf{r}_{\rm cm} = R_0 (\cos \theta_0, \, \sin \theta_0), \\
&\mathbf{v}_{\rm cm} = -v_0(\cos(\theta_0/2),\, \sin(\theta_0/2)),
\end{align}
where $\theta_0 = \arccos(2R_p/R_0-1)$ is the initial true anomaly and $v_0 = \sqrt{2GM_{\rm BH}/R_0}$ is the parabolic velocity.

We model a prograde, coplanar encounter where the internal configuration of the binary is determined by the initial orbital phase $\varphi_0$, defined relative to the initial radial direction.
The initial positions and velocities of the SLS ($m_1$) and the WD ($m_2$) are then:
\begin{align}
\mathbf{r}_{1,2} &= \mathbf{r}_{\rm cm}\pm \frac{a_0}{2} (\cos\varphi_0,\ \sin\varphi_0), \\
\mathbf{v}_{1,2} &= \mathbf{v}_{\rm cm}\pm \frac{v_b}{2} (-\sin\varphi_0, \cos\varphi_0),
\end{align}
where $v_b = \sqrt{Gm_b/a_0}$ is the relative orbital velocity.

While all other parameters remain fixed (see Table~\ref{tab:parameters}), the initial orbital phase $\varphi_0$ is the sole independent variable in our study. As shown in \cite{manzaneda2024relativistic}, the binary orientation at tidal interaction profoundly impacts the encounter's fate, with small variations in $\varphi_0$ leading to qualitatively different outcomes.

Two further orientation degrees of freedom are held fixed: the inclination $i$ between the binary plane and the plane of the CM trajectory, set to zero, and the sense of the binary's internal motion, taken to be prograde. A coplanar prograde encounter maximizes the coherence between the rotating binary and the rotating tidal field, and therefore the strength of phase effects, while collapsing the parameter space to $\varphi_0$ alone. We expect the trends reported below to weaken rather than change qualitatively as these assumptions are relaxed. For inclined encounters, only the projection of the binary separation vector onto the CM orbital plane couples efficiently to the tidal field, so the phase-dependent structure should narrow progressively with increasing $i$. For retrograde encounters, the binary rotates against the sense in which the tidal field sweeps around it, making the perturbation less coherent and the binary harder to separate \citep[e.g.,][]{hut1983topology}; deeper penetration would then be needed for comparable energy exchange. A systematic exploration of the $(\varphi_0, i)$ plane is deferred to future work.

\begin{table}
\centering
\caption{Summary of simulation parameters for binary-SMBH encounters.}
\label{tab:parameters}
\begin{tabular}{lll}
\hline
Parameter & Symbol & Value \\
\hline
\textbf{Black hole properties} & & \\
Mass & $M_{\rm BH}$ & $10^6 M_\odot$ \\
Schwarzschild radius & $R_{\rm Sch}$ & $4.24\,R\Sun$ \\
\hline
\textbf{Binary star properties} & & \\
SLS mass & $m_1$ & $1 M_\odot$ \\
SLS radius & $r_1$ & $1\,R\Sun$ \\
WD mass & $m_2$ & $1 M_\odot$ \\
WD radius & $r_2$ & $0.01\,R\Sun$ \\
Total binary mass & $m_b$ & $2 M_\odot$ \\
Initial separation & $a_0$ & $11\,R\Sun$ \\
Period & $T_b$ & 71.7\,h \\
\hline
\textbf{Tidal radii} & & \\
SLS tidal radius & $r_t$ & $100\,R\Sun$ \\
WD tidal radius & $r_{t,\rm WD}$ & $1\,R\Sun$ \\
Binary tidal radius & $R_t$ & $873\,R\Sun$ \\
\hline
\textbf{Orbital parameters} & & \\
Stellar impact parameter & $\beta_s$ & $1.2$ \\
Binary impact parameter & $\beta_b$ & $10.5$ \\
Pericenter distance & $r_p$ & $83.3\,R\Sun$ \\
Initial distance & $R_0$ & $30\,R_t$ \\
Orbit orientation & & Prograde, coplanar \\
\hline
\end{tabular}
\end{table}

\subsection{Preliminary three-body exploration}
\label{subsec:three_body}

We began by solving the restricted three-body problem for the binary--SMBH system, treating both stars as point masses. This allows us to efficiently explore the parameter space and identify the most dynamically interesting
configurations for detailed hydrodynamical study. Due to the equal masses of the binary components and the point-mass treatment, the interval $\varphi_0 \in [0^\circ, 180^\circ)$ produces outcomes identical to $\varphi_0 \in [180^\circ, 360^\circ)$ but with the roles of the SLS and WD exchanged. We therefore restrict our three-body exploration to $0^\circ < \varphi_0 < 180^\circ$.

\begin{figure}
    \centering
    \includegraphics[width=0.98\linewidth]{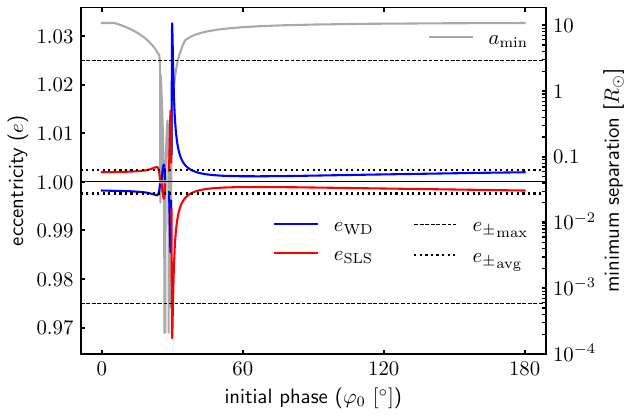}\\
    \includegraphics[width=0.98\linewidth]{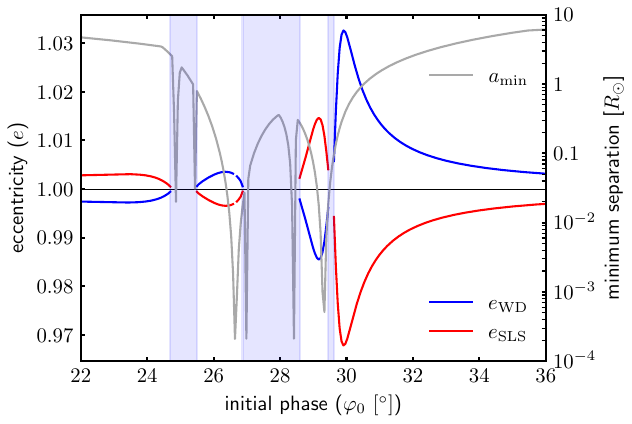}\\
    \caption{Outcomes of restricted three-body simulations showing binary encounters with a $10^6 M_\odot$ SMBH. \textbf{Top:} Final eccentricities of the bound (blue) and ejected (red) stars, and minimum internal SLS--WD separation reached during each encounter (gray), i.e.\ between the two stars themselves and not from the SMBH, across the full phase range $0^\circ$--$180^\circ$. Dashed and dotted lines show the analytic estimates for ${e_\pm}_\mathrm{max}$ and ${e_\pm}_\mathrm{avg}$ from equations~\eqref{eq:ecc_max} and~\eqref{eq:ecc_avg}, respectively. \textbf{Bottom:} Zoom into the critical phase range $22^\circ$--$36^\circ$, highlighting regions of binary survival (shaded) and phase-dependent identity of the ejected star. The most extreme eccentricities ($e = 1 \pm 0.033$) and stellar separations ($a_{\rm min} < 10^{-2} R\Sun$) occur near the survival-separation boundaries, indicating where hydrodynamic effects will be most significant.}
    \label{fig:3body}
\end{figure}

Figure~\ref{fig:3body} summarises the outcome of these simulations, showing the final eccentricities after binary separation and the minimum SLS--WD separation reached during each encounter, both as functions of $\varphi_0$. Throughout, $a_\mathrm{min}$ denotes the separation between the two stars themselves, not their distance from the SMBH. The top panel covers the full range ($0^\circ$--$180^\circ$), while the bottom panel provides a detailed view of the critical range $22^\circ < \varphi_0 < 36^\circ$. This last interval concentrates the full diversity of encounter outcomes: direct collisions, binary survival, alternating ejection identities, and the most extreme eccentricity deviations in the parameter space.

The resulting eccentricities exhibit a clear mirror symmetry about $e=1$, a direct consequence of energy conservation and the parabolic initial orbit of the binary CM. The mean eccentricity deviation across all phases is ${e_\pm}_\mathrm{avg} = 1 \pm 0.0021$, in excellent agreement with the analytic prediction of $1 \pm 0.0024$ from equation~\eqref{eq:ecc_avg}. The most significant deviations from parabolic orbits occur within a narrow phase interval around $\varphi_0 \simeq 30^\circ$, where the maximum deviation reaches $e_{\pm\mathrm{max}} = 1 \pm 0.033$, roughly consistent with the theoretical estimate $1 \pm 0.025$ from equation~\eqref{eq:ecc_max}. This confirms that optimal encounter geometries can produce substantially non-parabolic orbits, with important consequences for the resulting TDE signatures.

A key outcome of this exploration is the identification of a preliminary collisional window: the phase interval $24^\circ \lesssim \varphi_0 \lesssim 32^\circ$ within which the minimum separation satisfies $a_\mathrm{min}
\lesssim 1\,R_\odot$. Given the combined stellar radii ($R_\mathrm{SLS} + R_\mathrm{WD} = 1.01\,R_\odot$), these close approaches imply that direct physical collisions must occur in the corresponding hydrodynamical simulations.

Within the critical range, three distinct survival islands exist where the binary remains bound after the encounter: $(24.7^\circ, 25.5^\circ)$, $(26.9^\circ, 28.6^\circ)$, and $(29.4^\circ, 29.6^\circ)$. The identity of the ejected star alternates systematically across these islands, as evidenced by the transitions between the blue and red curves in the bottom panel of Figure~\ref{fig:3body}. The minimum separations within the window reach $a_\mathrm{min} < 10^{-2}\,R_\odot$ and as low as $\sim 10^{-4}\,R_\odot$ at specific phases, underscoring the extreme nature of these encounters.

\subsection{Hydrodynamical setup}
\label{subsec:hydro_setup}

We modelled the hydrodynamical evolution of the binary-SMBH encounters using the smoothed particle hydrodynamics (SPH) code \textsc{Phantom} \citep{price2018phantom}, which has been extensively validated for TDE studies \citep[e.g.,][]{bonnerot2016disc}. The SLS was modelled as a polytrope with $\gamma = 5/3$ and resolved with $5\times 10^5$ SPH particles. The WD was treated as a point mass, as justified in Section~\ref{subsec:parameters}.

Our restricted three-body simulations were initiated at a distance of 30\,$R_t$ from the SMBH to ensure accurate orbital initialization. To maintain computational tractability while preserving physical consistency in the hydrodynamical simulations, we initialized them using the system state extracted from the three-body integration at a distance of 5\,$R_t$. This is a direct hand-off rather than a fresh initialization: the positions and velocities at $5\,R_t$ are taken directly from the three-body integration begun at $R_0 = 30\,R_t$ with initial phase $\varphi_0$. The binary configuration is therefore inherited exactly, and $\varphi_0$ remains an unambiguous label for each run, the phase at the start of the SPH stage being $\varphi_0 + \omega_{\rm bin}\,\Delta t$, where $\omega_{\rm bin} = (G m_b / a_0^3)^{1/2}$ and $\Delta t$ is the time taken by the binary to travel between the two initial distances, obtained from Barker's equation (Section~\ref{subsec:hydro_simulations}).

While the restricted three-body dynamics exhibit a $180^\circ$ symmetry in $\varphi_0$, the hydrodynamical evolution does not. A shift of $\Delta \varphi_0 = 180^\circ$ maintains the binary's spatial orientation but interchanges the roles of the SLS and the WD. Since only the SLS is modelled as a deformable body, these two configurations represent physically distinct encounters (e.g., placing the SLS on the `internal' versus `external' side of the binary orbit at pericenter). We therefore explore the entire range $\varphi_0 \in (0, 360^\circ)$ to capture the full variety of hydrodynamical interactions.

To properly capture shock heating during the disruption process, we set the artificial viscosity parameters to $\alpha_{\rm AV} = 1.0$ and $\beta_{\rm AV} = 2.0$. These are the standard values recommended for \textsc{Phantom} \citep{price2018phantom} and used in previous SPH studies of TDEs \citep[e.g.,][]{bonnerot2016disc,mainetti2016hydrodynamical}. The parameters $\alpha_{\rm AV}$ and $\beta_{\rm AV}$ control the linear and quadratic terms of the artificial viscosity: the former acts as a bulk viscosity damping post-shock oscillations, while the latter dominates in strong, high Mach number shocks and prevents unphysical particle interpenetration. This is a numerical device for capturing shocks, not a physical viscosity (e.g., Shakura--Sunyaev), and it is inactive away from converging flows. Gravitational forces were computed using a binary tree with an opening angle $\theta = 0.5$, and the simulations employed individual particle timesteps with a Courant factor of 0.3.

Each simulation was evolved for approximately 100 hours around pericenter passage ($\sim50$ hours before to $\sim50$ hours after), with $t=0$ defined as the moment of pericenter passage. This duration is sufficient to capture the stabilization of the debris energy distribution in most of the encounters (see discussion later on). By ``stabilization'' we mean that the specific orbital energy of the gas has stopped evolving, so that the energy distribution, and hence the predicted fallback rate, is frozen. This is distinct from circularization, which is not expected until the debris returns to pericenter, long after our runs end, and which we do not follow. Operationally, we regard the debris as stabilized once the bound mass fraction of the SLS gas changes by less than $0.02$ per cent between consecutive outputs. This criterion motivated our choice of $t = 53$ hours as the reference time for the analysis presented in this work (see Appendix~\ref{app:time_conv} for details).

We performed extensive convergence tests to verify the reliability of our results. As detailed in Appendix~\ref{app:convergence}, our standard resolution of $5\times 10^5$ particles yields converged results for the bulk dynamics on the simulated timescale. Additionally, tests varying the initial binary-SMBH separation between $1$--$10$ $R_t$ showed no significant differences in the outcomes, validating our choice of initial conditions.

\subsection{Hydrodynamical simulations}
\label{subsec:hydro_simulations}

Guided by the three-body parameter-space exploration of Section~\ref{subsec:three_body}, we performed a suite of 48 hydrodynamical simulations in which the initial binary phase $\varphi_0$ is the only free parameter, enabling a systematic investigation of the full spectrum of disruption outcomes. The sample covers the whole $360^\circ$ range of $\varphi_0$, with higher resolution within the collisional windows. The initial parameters and main characteristics of all simulations are summarised in Table~\ref{tab:sim_summary}.

In particular, the phase columns report both the initial binary phase $\varphi_0$ and the ideal binary phase at the tidal radius, $\varphi^{*}_{R_t}$, which represents the phase the system would have at $R_t$ in the absence of tidal forces. To compute $\varphi^{*}_{R_t}$, we estimate the travel time $\Delta t$ from $30R_t$ to $R_t$ by solving Barker's equation for parabolic motion \citep[e.g.,][]{galatic}, resulting in $\Delta t \simeq 36.8$ days for our fiducial parameters. The phase is then given by $\varphi^{*}_{R_t} = \varphi_0 + \omega_{\rm bin}\Delta t$.

This distinction matters for any future use of our results. The label $\varphi_0$ is tied to our choice $R_0 = 30\,R_t$: a different $R_0$ would relabel the same physical encounters by a constant offset $\omega_{\rm bin}\Delta t$, shifting the quoted boundaries of the collisional window. The $R_0$-independent quantity is $\varphi^{*}_{R_t}$, in terms of which the collisional windows identified below correspond to $\varphi^{*}_{R_t} \simeq 134.7^\circ$--$141.2^\circ$ and $314.7^\circ$--$321.2^\circ$. We recommend that comparisons with our results be made in this variable.

We further include a single-star benchmark simulation (SBM) in which the SLS follows an orbit identical to that of the binary CM. This control case establishes the baseline fallback rate behaviour of a classical parabolic TDE, against which all binary-driven encounters are compared throughout this work.

\begin{table*}
\centering
\caption{Summary of binary--SMBH encounter simulations. Simulation names encode the initial binary phase $\varphi_0$ in degrees. Columns show: simulation name; initial phase of the binary system; phase of the binary system at tidal radius; predicted eccentricities from three-body dynamics; resulting eccentricities from hydrodynamical simulations; TDE classification; the mass of the resulting White Dwarf with Debris Envelope (WDDE); and its characteristic radius. TDE types are: Elliptical TDE (eTDE), Hyperbolic TDE (hTDE), Collision-driven TDE (cTDE). All simulations use $\beta_{s}=1.2$ ($\beta_{b}=10.5$).}
\label{tab:sim_summary}
\begin{tabular}{lccccccccc}
\hline
Simulation &
\multicolumn{2}{c}{Phase ($\varphi$)} &
\multicolumn{2}{c}{Three-body} &
\multicolumn{2}{c}{Hydrodynamical} &
TDE type & WD/WDDE mass & WD/WDDE radius \\
\cline{2-7}
 & Initial ($\varphi_0$) & Ideal ($\varphi^{*}_{R_t}$) & WD $e$ & SLS $e$ & WD $e$ & SLS $e$ & & [$M_\odot$] & [$R_\odot$] \\
\hline
S20 & $20.0^\circ$ & $128.65^\circ$ & 0.9976 & 1.0025 & 0.9973 & 1.0039 & hTDE & 1.09 & 27.9620 \\
S22 & $22.0^\circ$ & $130.65^\circ$ & 0.9974 & 1.0028 & 0.9969 & 1.0055 & hTDE & 1.1501 & 28.5842 \\
S24 & $24.0^\circ$ & $132.65^\circ$ & 0.9975 & 1.0027 & 0.9965 & 1.0083 & hTDE & 1.2463 & 23.2175 \\
S25 & $25.0^\circ$ & $133.65^\circ$ & 1.0002 & 0.9999 & 0.9963 & 1.0100 & hTDE & 1.2896 & 20.5210 \\
S26 & $26.0^\circ$ & $134.65^\circ$ & 1.0029 & 0.9972 & 0.9961 & 1.0122 & cTDE & 1.3565 & 12.7517 \\
S27 & $27.0^\circ$ & $135.65^\circ$ & \multicolumn{2}{c}{binary survives} & 0.9963 & 1.0117 & cTDE & 1.4334 & 9.2314 \\
S28 & $28.0^\circ$ & $136.65^\circ$ & \multicolumn{2}{c}{binary survives} & 0.9991 & 1.0069 & cTDE & 1.5976 & 8.5977 \\
S28.5 & $28.5^\circ$& $137.15^\circ$ & \multicolumn{2}{c}{binary survives} & 1.0010 & 0.9993 & cTDE & 1.6047 & 5.4167 \\
S29 & $29.0^\circ$ & $137.65^\circ$ & 0.9878 & 1.0123 & 1.0051 & 0.9804 & cTDE & 1.4937 & 5.6348 \\
S29.8 & $29.8^\circ$& $138.45^\circ$& 1.0292 & 0.9710 & 1.0121 & 0.9726 & cTDE & 1.2854 & 4.8936 \\
S30 & $30.0^\circ$& $138.65^\circ$ & 1.0317 & 0.9686 & 1.0136 & 0.9743 & cTDE & 1.2278 & 4.6521 \\
S30.2 & $30.2^\circ$& $138.85^\circ$ & 1.0267 & 0.9736 & 1.0144 & 0.9766 & cTDE & 1.1738 & 4.4238 \\
S30.5 & $30.5^\circ$& $139.15^\circ$ & 1.0200 & 0.9801 & 1.0143 & 0.9800 & cTDE & 1.1239 & 4.3122 \\
S31 & $31.0^\circ$& $139.65^\circ$ & 1.0136 & 0.9865 & 1.0129 & 0.9846 & cTDE & 1.0854 & 4.3227 \\
S32.5 & $32.5^\circ$& $141.15^\circ$ & 1.0068 & 0.9933 & 1.0075 & 0.9920 & cTDE & 1.0792 & 9.8145 \\
S33 & $33.0^\circ$& $141.65^\circ$ & 1.0058 & 0.9943 & 1.0064 & 0.9929 & eTDE & 1.1006$^{\star}$ & 10.0$^{\star}$ \\
S35 & $35.0^\circ$& $143.65^\circ$ & 1.0037 & 0.9964 & 1.0044 & 0.9940 & eTDE & 1.1247$^{\star}$ & 10.0$^{\star}$ \\
S37 & $37.0^\circ$& $145.65^\circ$ & 1.0027 & 0.9973 & 1.0036 & 0.9949 & eTDE & 1.0699$^{\star}$ & 10.0$^{\star}$ \\
S40 & $40.0^\circ$& $148.65^\circ$ & 1.0020 & 0.9980 & 1.0023 & 0.9971 & eTDE & 1.0001 & 0.01 \\
S70 & $70.0^\circ$ & $178.65^\circ$ & 1.0011 & 0.9989 & 1.0011 & 0.9988 & eTDE & 1.0 & 0.01 \\
S90 & $90.0^\circ$ & $198.65^\circ$ & 1.0013 & 0.9988 & 1.0012 & 0.9987 & eTDE & 1.0 & 0.01 \\
S100 & $100.0^\circ$ & $208.65^\circ$ & 1.0013 & 0.9987 & 1.0013 & 0.9987 & eTDE & 1.0 & 0.01 \\
S120 & $120.0^\circ$ & $228.65^\circ$ & 1.0015 & 0.9986 & 1.0015 & 0.9985 & eTDE & 1.0 & 0.01 \\
S150 & $150.0^\circ$ & $258.65^\circ$ & 1.0018 & 0.9984 & 1.0017 & 0.9983 & eTDE & 1.0 & 0.01 \\
\hline
S200 & $200.0^\circ$& $308.65^\circ$ & 1.0025 & 0.9976 & 1.0028 & 0.9959 & eTDE & 1.095 & 28.3875 \\
S202 & $202.0^\circ$& $310.65^\circ$ & 1.0028 & 0.9974 & 1.0032 & 0.9944 & eTDE & 1.1591 & 29.0353 \\
S204 & $204.0^\circ$& $312.65^\circ$ & 1.0027 & 0.9975 & 1.0034 & 0.9917 & eTDE & 1.2507 & 23.6071 \\
S205 & $205.0^\circ$& $313.65^\circ$ & 0.9999 & 1.0002 & 1.0038 & 0.9903 & eTDE & 1.2928 & 20.7392 \\
S206 & $206.0^\circ$& $314.65^\circ$ & 0.9972 & 1.0029 & 1.0040 & 0.9882 & cTDE & 1.3597 & 12.9680 \\
S207 & $207.0^\circ$& $315.65^\circ$ & \multicolumn{2}{c}{binary survives} & 1.0037 & 0.9887 & cTDE & 1.4295 & 9.9447 \\
S208 & $208.0^\circ$& $316.65^\circ$ & \multicolumn{2}{c}{binary survives} & 1.0013 & 0.9938 & cTDE & 1.5581 & 10.1148 \\
S208.5 & $208.5^\circ$& $317.15^\circ$ & \multicolumn{2}{c}{binary survives} & 0.9988 & 1.0017 & cTDE & 1.5961 & 5.3450 \\
S209 & $209.0^\circ$& $317.65^\circ$ & 1.0123 & 0.9878 & 0.9950 & 1.0188 & cTDE & 1.4955 & 5.5877 \\
S209.8 & $209.8^\circ$& $318.45^\circ$ & 0.9710 & 1.0292 & 0.9878 & 1.0274 & cTDE & 1.2771 & 4.4958 \\
S210 & $210.0^\circ$& $318.65^\circ$ & 0.9686 & 1.0317 & 0.9865 & 1.0260 & cTDE & 1.2266 & 4.6304 \\
S210.2 & $210.2^\circ$& $318.85^\circ$ & 0.9736 & 1.0267 & 0.9857 & 1.0239 & cTDE & 1.1798 & 4.0713 \\
S210.5 & $210.5^\circ$& $319.15^\circ$ & 0.9801 & 1.0200 & 0.9857 & 1.0207 & cTDE & 1.1320 & 4.3213 \\
S211 & $211.0^\circ$& $319.65^\circ$ & 0.9865 & 1.0136 & 0.9872 & 1.0157 & cTDE & 1.0869 & 4.2609 \\
S212.5 & $212.5^\circ$& $321.15^\circ$ & 0.9933 & 1.0068 & 0.9925 & 1.0081 & cTDE & 1.0774 & 9.3199 \\
S213 & $213.0^\circ$& $321.65^\circ$ & 0.9943 & 1.0058 & 0.9937 & 1.0071 & hTDE & 1.1027$^{\star}$ & 10.0$^{\star}$ \\
S215 & $215.0^\circ$& $323.65^\circ$ & 0.9964 & 1.0037 & 0.9956 & 1.0060 & hTDE & 1.1298$^{\star}$ & 10.0$^{\star}$ \\
S217 & $217.0^\circ$& $325.65^\circ$ & 0.9973 & 1.0027 & 0.9964 & 1.0051 & hTDE & 1.0584$^{\star}$ & 10.0$^{\star}$ \\
S220 & $220.0^\circ$& $328.65^\circ$ & 0.9980 & 1.0020 & 0.9977 & 1.0027 & hTDE & 1.0001 & 0.01 \\
S250 & $250.0^\circ$ & $358.65^\circ$ & 0.9989 & 1.0011 & 0.9989 & 1.0010 & hTDE & 1.0 & 0.01 \\
S270 & $270.0^\circ$ & $18.65^\circ$ & 0.9988 & 1.0013 & 0.9988 & 1.0012 & hTDE & 1.0 & 0.01 \\
S280 & $280.0^\circ$ & $28.65^\circ$ & 0.9987 & 1.0013 & 0.9987 & 1.0012 & hTDE & 1.0 & 0.01 \\
S300 & $300.0^\circ$ & $48.65^\circ$ & 0.9986 & 1.0015 & 0.9986 & 1.0014 & hTDE & 1.0 & 0.01 \\
S330 & $330.0^\circ$ & $78.65^\circ$ & 0.9984 & 1.0018 & 0.9984 & 1.0017 & hTDE & 1.0 & 0.01 \\
\hline
\end{tabular}
\begin{tablenotes}
      \item $^{\star}$Identifies systems with a bound debris envelope that remains spatially extended and has not yet settled into a stable, spherical configuration. The envelope radius, and hence any quantity derived from it, should be read as an upper limit for these systems.
\end{tablenotes}
\end{table*}

\section{Hydrodynamical evolution and outcomes}
\label{sec:dynamics}

This section describes the hydrodynamical evolution of the binary--SMBH encounter, focusing on how the stellar interaction determines the final gas morphology and subsequent fallback rate dynamics. While the restricted three-body problem provides a useful first approximation, accounting for hydrodynamics introduces qualitatively new phenomena: tidal distortion, mass stripping, and, in the most extreme cases, direct physical collisions.

\subsection{Global evolution}
\label{subsec:global_evolution}

\begin{figure*}
    \centering
    \includegraphics[width=17.5cm]{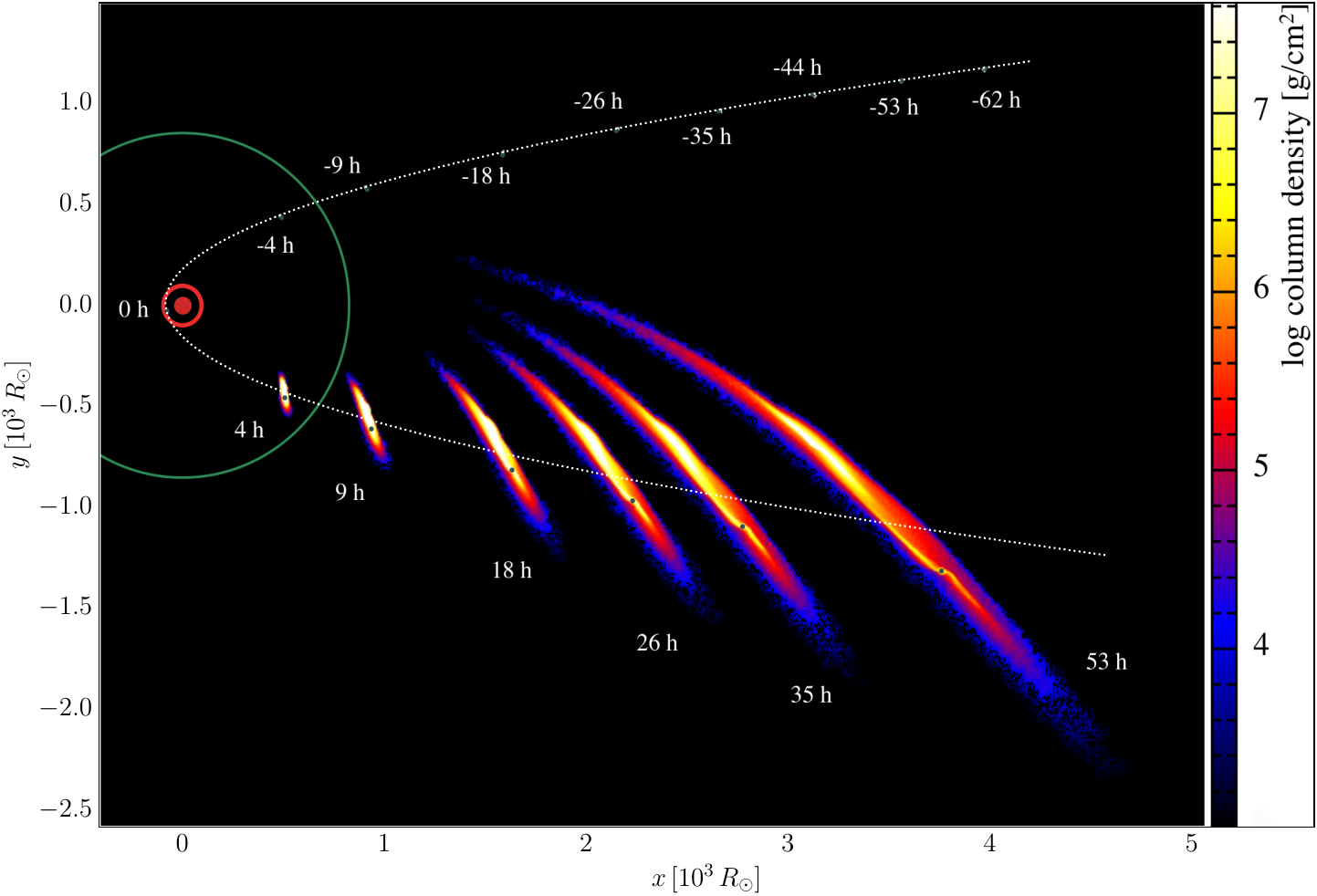}
    \caption{Column density maps of simulation S30 at key moments during the encounter with the SMBH (located at the origin). The white dashed line traces the trajectory of the binary's CM. The red and green circles indicate the tidal radius of the SLS ($r_t = 100\,\mathrm{R_\odot}$) and the binary system ($R_t = 873\,\mathrm{R_\odot}$), respectively. The grey point at each time marks the position of the WD. Times are given relative to pericenter passage ($t=0$). Near pericenter, as the SLS is being disrupted, a collision with the WD occurs. By the final snapshot ($t \simeq 53\,\mathrm{h}$), the debris has expanded to thousands of solar radii and reached an energetically stable configuration, setting the stage for the fallback rate phase. Rendered using \texttt{SPLASH} \citep{2007SPLASH}.}
    \label{fig:simulacion_30}
\end{figure*}

To illustrate the global dynamical evolution we show in Figure~\ref{fig:simulacion_30} the case S30 ($\varphi_0 = 30^\circ$) as a representative example of a collisional encounter. The figure presents column-density maps at different times together with the trajectory of the CM of the binary system, indicated by the white dotted line. We define $t=0$\,h as the moment of closest approach to the SMBH.

Following the interaction, the SLS debris expands rapidly (visible for $t > 4\,\mathrm{h}$). Depending on the encounter geometry, the WD may capture a fraction of this disrupted material and become embedded within the debris stream. By $t = 53\,\mathrm{h}$ the debris has expanded to thousands of solar radii; most of it has reached an energetically stable configuration, although a persistent concentration remains gravitationally bound to the WD, forming an extended non-degenerate envelope that is still relaxing at this time. We refer to this composite object hereafter as a White Dwarf with a Debris Envelope (WDDE). From this point onward, the bound material progressively falls back toward the SMBH while the unbound component recedes indefinitely.

\subsection{Morphological diversity}
\label{subsec:morphological_diversity}

While S30 illustrates the most extreme class of encounter, the outcome varies dramatically with the initial binary phase $\varphi_0$. As established in Section~\ref{subsec:three_body}, the collisional window spans $\varphi_0 \approx 24^\circ$--$32^\circ$ (and $204^\circ$--$212^\circ$); we use this classification from the outset to organise the morphological survey that follows.

\begin{figure*}
    \includegraphics[width=17.5cm]{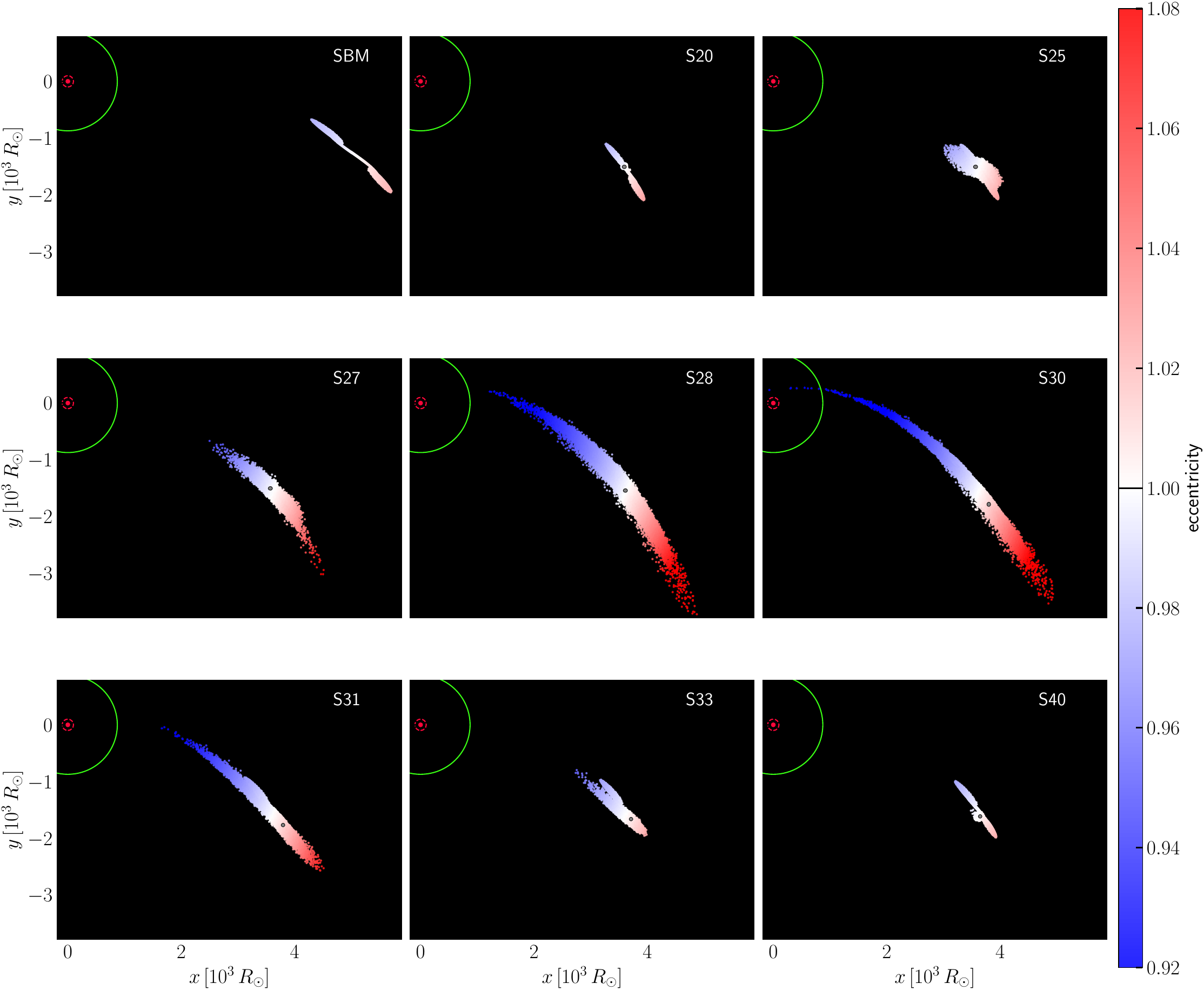}
    \caption{
Eccentricity maps of the gas $\simeq 53\,\mathrm{h}$ after pericenter passage for different initial orbital phases of the binary system. The upper-left panel shows the benchmark case of a single star on a parabolic orbit (SBM). In the remaining panels, the grey circle marks the position of the WD. Collisions occurring near pericenter ($25^\circ$--\,$33^\circ$) produce a more disturbed morphology, a faster return of the material, and a wider range of eccentricities, while the extreme cases ($20^\circ$ and $40^\circ$) retain eccentricity distributions closer to $e \sim 1$, similar to the classical scenario. The red and green circles indicate the tidal radius of the SLS ($r_t = 100\,\mathrm{R_\odot}$) and the binary system ($R_t = 873\,\mathrm{R_\odot}$), respectively.}
\label{fig:ecc_maps}
\end{figure*}

Figure~\ref{fig:ecc_maps} captures this diversity through eccentricity maps at $t \simeq 53\,\mathrm{h}$ for nine simulations spanning $\varphi_0 = 20^\circ$\,--\,$40^\circ$, together with the single-star benchmark SBM. The single-star case SBM (top left panel) exhibits the classical parabolic disruption morphology: a thin, nearly symmetric stream with eccentricities narrowly concentrated around $e = 1$ ($e \simeq 1 \pm 0.02$) \citep[e.g.,][]{lodato2009stellar,guillochon2013hydrodynamical, Stone2013}. For S20 the morphology remains broadly similar, consistent with a non-collisional encounter. A dramatic transition occurs across S25--S33, where the stream appears broader and increasingly asymmetric, with the eccentricity distribution widening up to $e \simeq 1 \pm 0.1$. In the most extreme cases (e.g.\ S28, S30, and S31), the bound component is highly elongated and shows signs of already returning toward the SMBH by the final time, reflecting the profound redistribution of orbital energies caused by the direct WD--SLS collision. By S40, the morphology again resembles SBM, although the now-unbound WD has interacted with the outer debris layers,
imprinting subtle asymmetries in the distribution.

\begin{figure*}
    \centering
    \includegraphics[width=17.5cm]{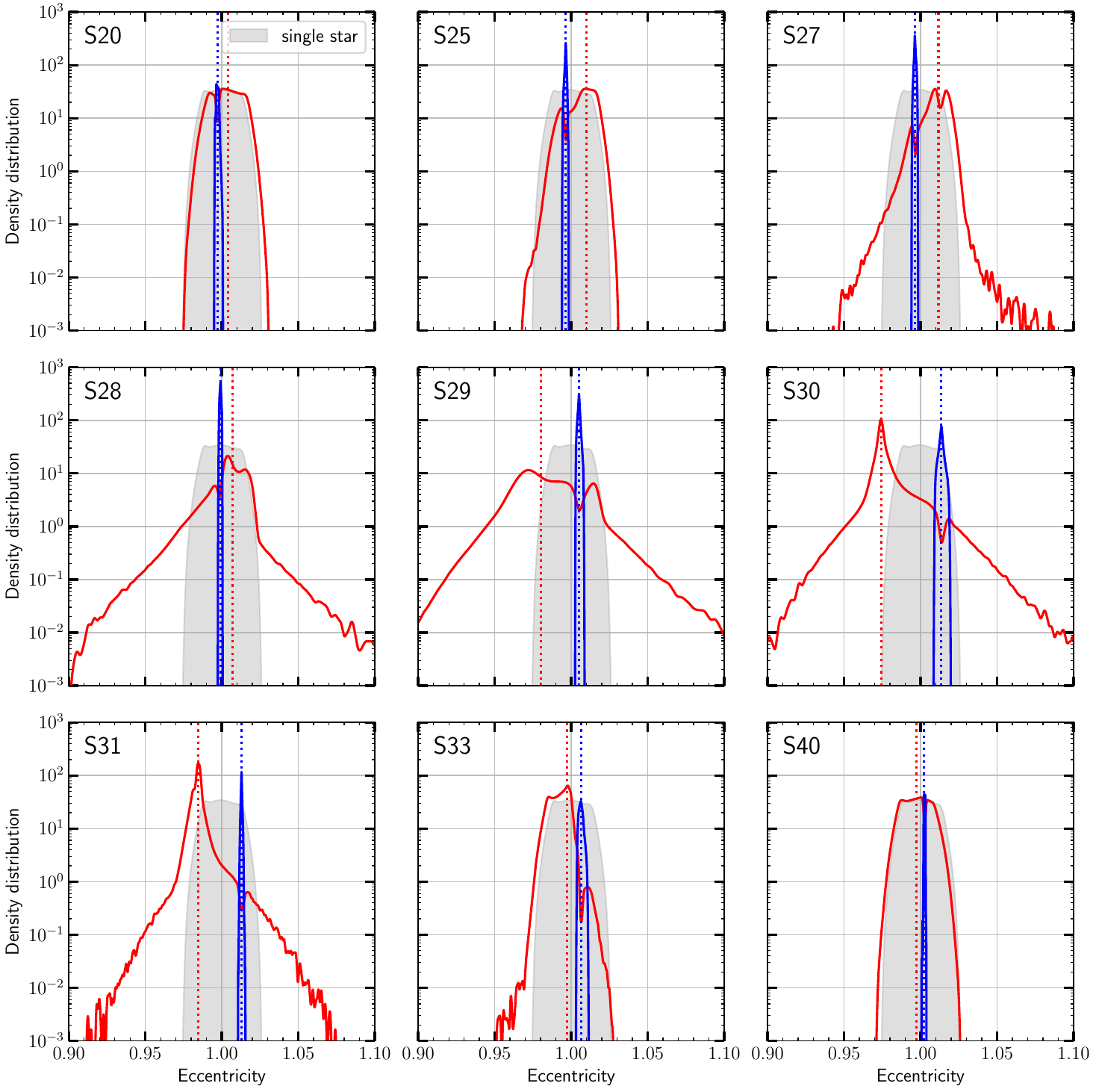}
    \caption{Eccentricity distributions of the gas at $t = 53\,\mathrm{h}$ for simulations S20--S40. Red shows the disrupted SLS debris; blue shows material gravitationally bound to the WD (WDDE). The shaded region shows the single-star benchmark SBM for reference. The gap in the red distribution coincides with the narrow blue peak, reflecting debris swept up by the WD and subsequently relaxing into the WDDE. Vertical dotted lines indicate the eccentricity of the WD (blue) and the mean eccentricity value of the SLS debris (red).}
    \label{fig:histograms_all}
\end{figure*}

Figure~\ref{fig:histograms_all} complements the spatial picture by showing the final eccentricity distributions of the gas for the same simulation range. Red shows the disrupted SLS debris, blue the material gravitationally bound to the WD (WDDE), identified by the criterion described in Appendix~\ref{app:bound_mass_wd}, and the shaded region the single-star benchmark SBM.
To obtain smooth eccentricity distributions and reduce the noise associated with finite particle sampling, we employed a Gaussian kernel density estimator (KDE), following the approach of \citet{addison2019busting}. This non-parametric method reconstructs the underlying probability density function by replacing each particle with a Gaussian kernel of finite width and summing the contribution of all particles.

For non-collisional encounters (e.g., S20 and S40), the red distribution closely resembles SBM in overall shape. In these cases, a well-defined gap in the red distribution coincides with the narrow blue peak of the WDDE, directly reflecting the fraction of debris swept up by the WD. In collisional encounters (S27--S33), by contrast, the red distribution broadens substantially, departing from SBM in both shape and extent. A particularly notable feature appears in S30 and S31, where the red distribution exhibits a distinct peak. This is the signature of a surviving SLS core that has withstood complete tidal disruption.

\begin{figure}
    \centering

    \begin{overpic}[width=\linewidth]{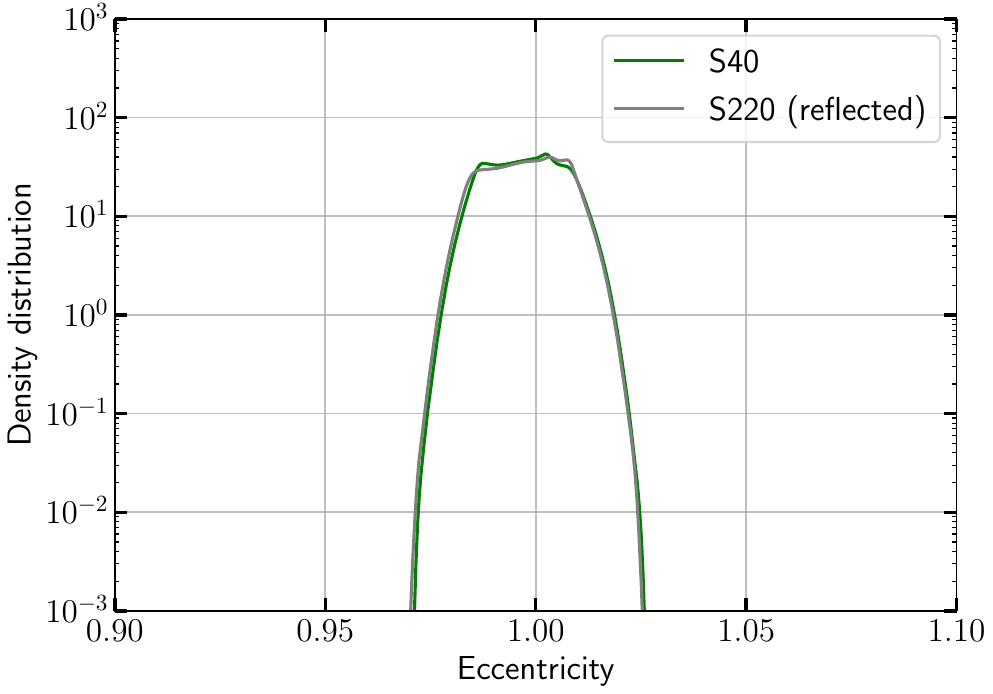}
    \put(120,135){WDDE}
    \end{overpic}\\

    \begin{overpic}[width=\linewidth]{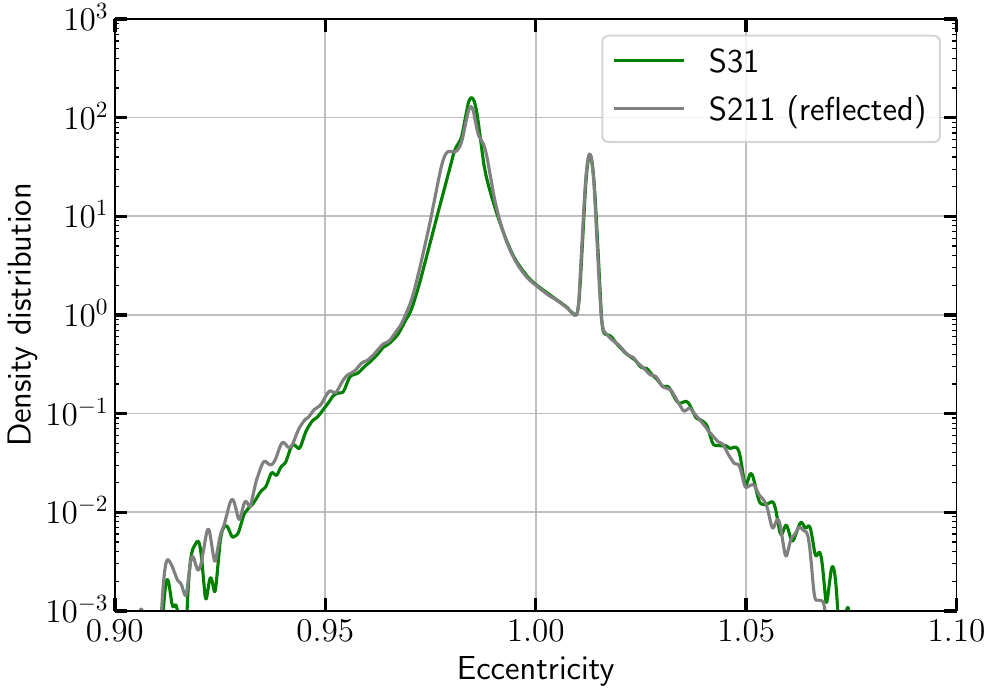}
        \put(95,150){surviving core}
        \put(150,120){WDDE}
    \end{overpic}\\

    \begin{overpic}[width=\linewidth]{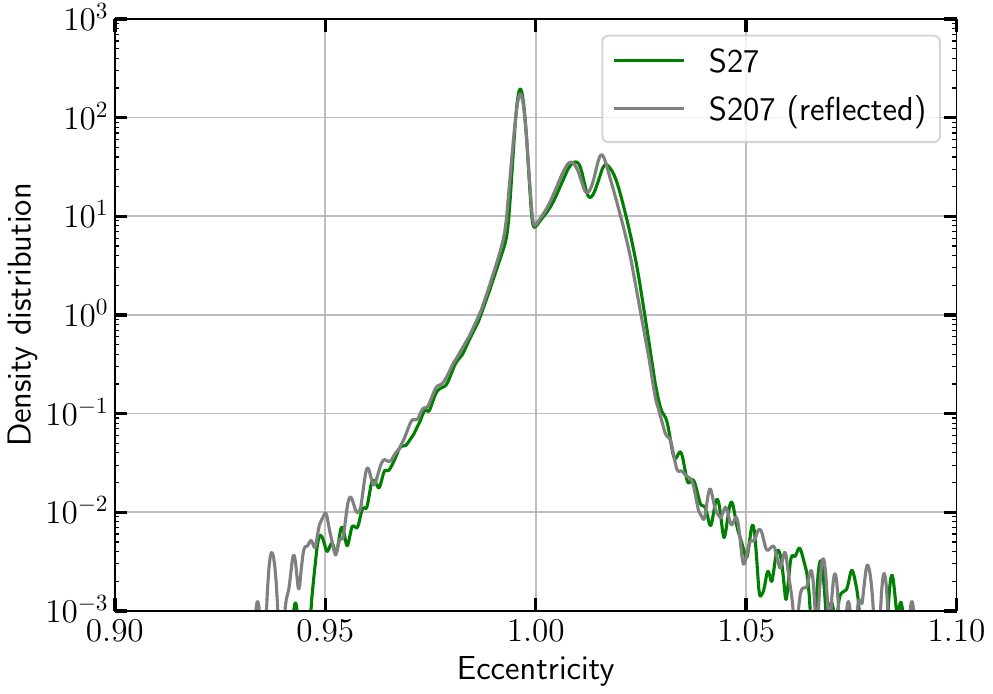}
        \put(100,145){WDDE}
    \end{overpic}

    \caption{Eccentricity distributions of the disrupted SLS debris for paired simulations in the S20--S40 (green) and S200--S220 (grey) ranges, which differ only in the initial binary phase by $180^\circ$. From top to bottom: S40 vs.\ S220, S31 vs.\ S211, and S27 vs.\ S207. The grey distributions are inverted about the horizontal axis to facilitate direct comparison. Despite producing opposite bound/unbound partitions for the WD, the two ranges yield broadly equivalent distributions. }
    \label{fig:reflejado}
\end{figure}

The complementary range S200--S220 exhibits a nearly identical morphological progression, albeit with the bound/unbound partition effectively reversed. In this sequence, the WD is unbound for S200 and bound for S220, mirroring the dynamical behaviour observed in the S20--S40 range. This equivalence is clearly illustrated in Figure~\ref{fig:reflejado}, which compares the eccentricity distributions of the paired simulations S27 vs.\ S207, S31 vs.\ S211, and S40 vs.\ S220. The grey distributions (S200--S220), inverted about the horizontal axis to facilitate direct comparison, closely mirror their green counterparts (S20--S40) in both shape and width, confirming that the two ranges are energetically equivalent despite interchanging the final bound/unbound identity of the WD.

\subsection{The role of stellar collisions}
\label{subsec:collisions}

\begin{figure}
    \centering
    \includegraphics[width=\linewidth]{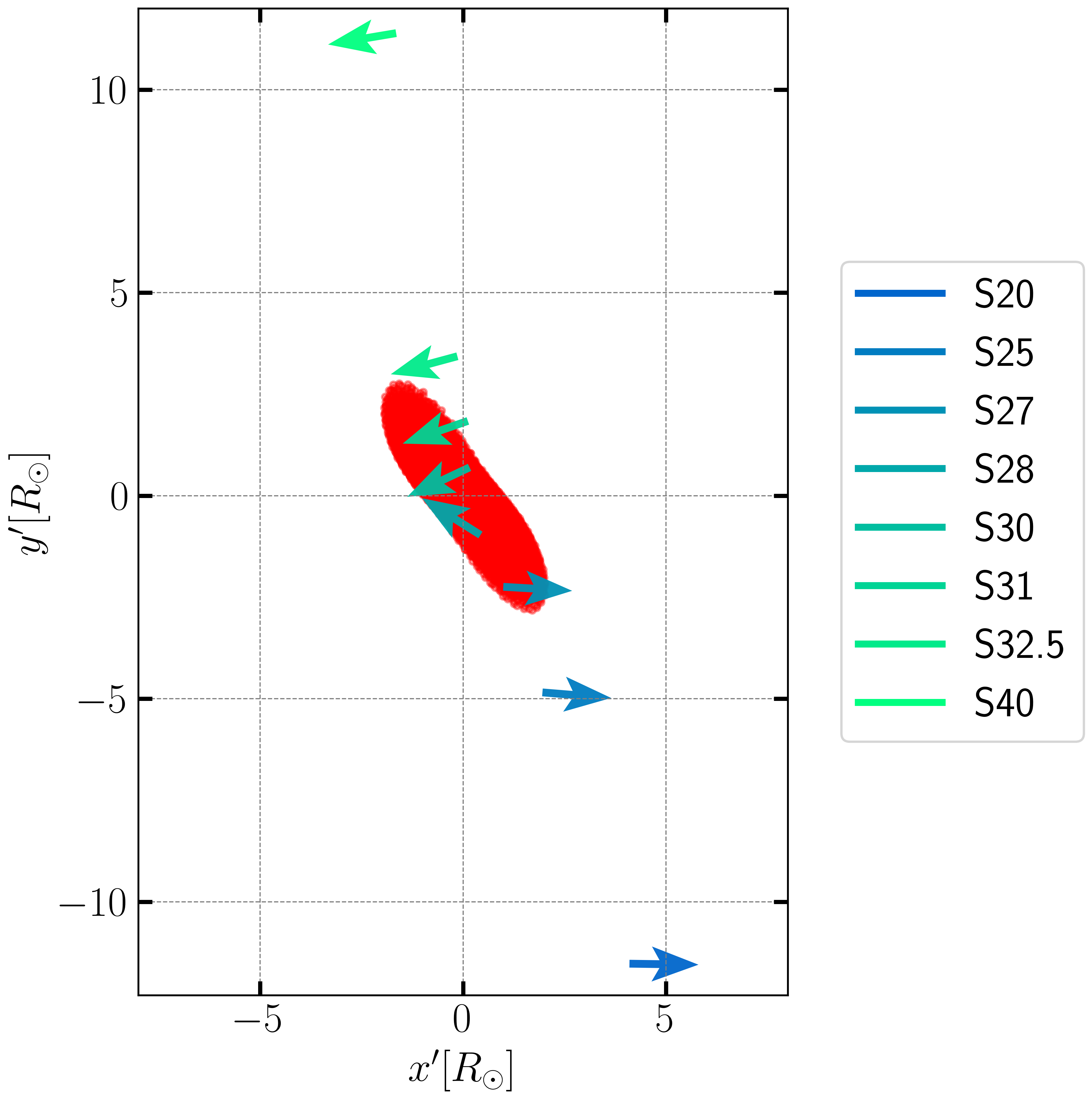}
    \caption{Geometry of the WD--SLS encounter near pericenter for simulations S20--S40, in the SLS centre-of-mass frame. The tidally distorted SLS (representative shape taken from simulation S32.5) is centred at the origin. Each marker indicates the WD position at closest approach, with the arrow showing the direction of relative motion of the WD at that instant; markers inside the stellar boundary identify collisional encounters, while those outside correspond to flyby interactions. The configurations corresponding to S200--S220 follow the same geometry but with the opposite sense of motion.}
    \label{fig:collision_comparison}
\end{figure}

The morphological transition visible in Figures~\ref{fig:ecc_maps} and \ref{fig:histograms_all} points to a single physical cause: whether or not the WD collides directly with the SLS near pericentre. This is made explicit in Figure~\ref{fig:collision_comparison}, which shows the geometry of the WD--SLS encounter near pericenter in the SLS centre-of-mass frame. The tidally distorted SLS is centred at the origin, with WD positions shown for simulations S20--S40. The corresponding configurations for S200--S220 follow the same geometry but with the opposite sense of motion. The figure makes immediately apparent whether the WD penetrates the SLS or passes at a safe distance, confirming that the collisional window identified from three-body calculations in Section~\ref{subsec:three_body} persists in the full hydrodynamical simulations. Within this window, the direct WD--SLS collision redistributes orbital energy between the stars, broadens the debris eccentricity distribution, and enables substantial mass capture by the WD, as quantified in Section~\ref{sec:WDDE}.

In non-collisional encounters, by contrast, the WD undergoes only a milder interaction with the expanding debris after pericentre passage. Depending on the encounter geometry, this interaction can nonetheless leave a measurable imprint on the fallback rate, as we discuss in Section~\ref{sec:menagerie}.

\subsection{Comparison with three-body dynamics}
\label{subsec:comparison_3body}

\begin{figure}
\centering
\begin{minipage}{0.49\columnwidth}
    \centering
    \includegraphics[width=\linewidth]{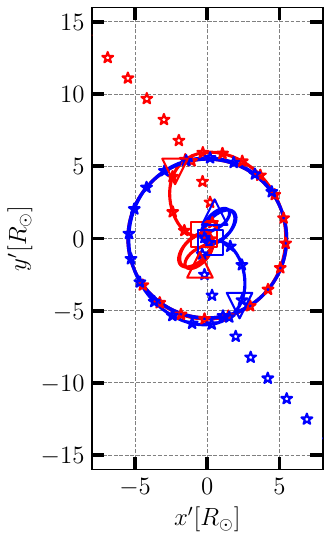}
\end{minipage}
\hfill
\begin{minipage}{0.49\columnwidth}
    \centering
    \includegraphics[width=\linewidth]{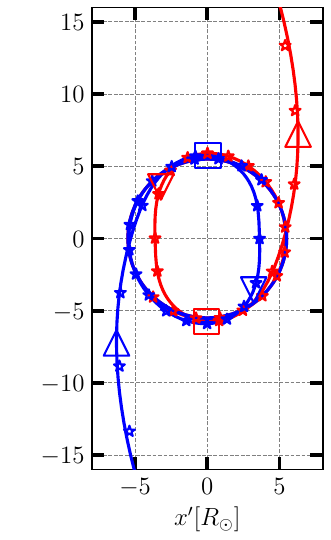}
\end{minipage}
\caption{Trajectories of the binary system in the comoving reference frame of the binary CM. Solid lines correspond to three-body simulations, while `$\star$' markers represent hydrodynamical simulations. Squares indicate pericenter passage, downward-pointing triangles mark entry into the tidal radius, and upward-pointing triangles denote exit from this region. The left panel shows the S28.5 case, where the two approaches diverge significantly, while the right panel corresponds to the S40 case, where both trajectories are largely consistent.}
\label{fig:trajectories}
\end{figure}

The restricted three-body calculations of Section~\ref{subsec:three_body} treat both stars as point masses and thereby provide a clean baseline against which to assess hydrodynamical effects.
Figure~\ref{fig:trajectories} illustrates this
comparison directly in the reference frame of the binary CM for two representative
cases. The right panel shows S40, a non-collisional encounter: the SPH
trajectories of the WD and the SLS centre of mass agree remarkably well with
the three-body predictions throughout the entire evolution, confirming that
point-mass dynamics captures the essential physics when the stars remain well
separated. The left panel shows S28.5, a collisional encounter: the two
approaches agree closely for $t < 0$, but diverge drastically after pericenter
passage, when the physical collision fundamentally alters the subsequent
evolution in ways that the point-mass treatment cannot capture.

\begin{figure}
    \centering
    \includegraphics[width=\linewidth]{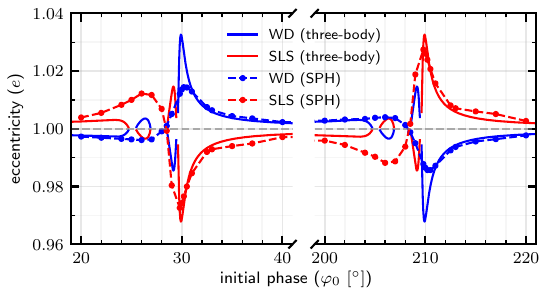}
    \caption{Final eccentricity of the WD and mean eccentricity of the SLS debris as a function of initial binary phase $\varphi_0$, comparing three-body integrations (solid lines) with SPH simulations (points). Bound and unbound regimes are separated by the dashed line at $e = 1$.}
    \label{fig:ecc_3body_hydro}
\end{figure}

Figure~\ref{fig:ecc_3body_hydro} extends this comparison to the final eccentricities of all simulations as a function of $\varphi_0$. Outside the collisional window, the agreement is remarkably good, with SPH eccentricities tracking the three-body predictions to within $\Delta e \lesssim 0.005$. The bound-to-unbound transition of the WD, occurring at $\varphi_0 \approx 28.4^\circ$ and $208.4^\circ$ in both approaches, further supports this picture, indicating that the energy exchange governing which star remains bound is largely set by gravitational dynamics prior to any collision.
Within the collisional windows, however, significant departures emerge. Most notably, the three-body calculations predict three distinct survival islands where the binary remains bound after the encounter (shaded regions in the bottom panel of Figure~\ref{fig:3body}). In the SPH simulations, these survival islands disappear entirely, as the physical collisions and tidal disruption that occur in this regime inevitably unbind the binary.

An additional regularity, preserved in both approaches, is the near-equal and inverse partitioning of outcomes: for roughly half of all initial phases the WD remains bound to the SMBH while the bulk of the SLS debris is ejected, and vice versa. This complementarity arises from both energy conservation and the parabolic centre-of-mass trajectory.

While the three-body approach provides an excellent first approximation and correctly predicts the gross partitioning of outcomes, it breaks down precisely where hydrodynamics matters most, in the collisional regime responsible for the most exotic TDEs and the most significant mass capture by the WD.

Further details on the comparison between the hydrodynamical simulations and the three-body predictions are given in Appendix~\ref{app:comparison_3body}, where we quantify the deviations in the orbital trajectories. These deviations are present both within the collision window and in encounters where no direct collision occurs, although they are significantly enhanced in the former due to strong interactions at pericenter. In the latter case, the trajectories still exhibit moderate deviations as a result of the tidal disruption of the SLS.

\section{Mass budget and the fate of the WDDE}
\label{sec:WDDE}

The hydrodynamical simulations described in Section~\ref{sec:dynamics} produce
a wide variety of post-encounter configurations, differing not only in the
spatial morphology of the debris but also in the amount of mass partitioned
among the three possible fates: accretion onto the SMBH, capture by the WD,
and ejection from the system. Understanding this mass partitioning is essential
both for characterising the long-term evolution of the WDDE and for
interpreting the diversity of fallback rate curves presented in
Section~\ref{sec:menagerie}.

\subsection{Mass budget}
\label{subsec:mass_budget}

\begin{figure*}
    \centering
    \includegraphics[width=\linewidth]{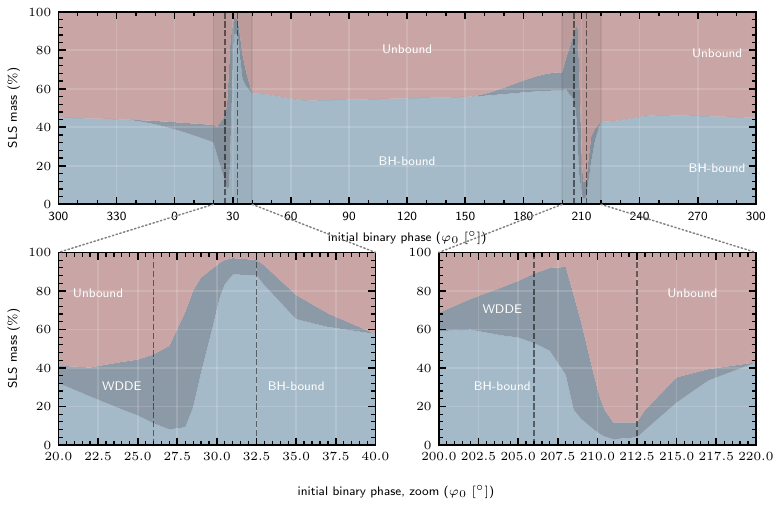}
    \caption{Mass budget at $t = 53\,\mathrm{h}$ as a function of initial binary
phase $\varphi_0$ for all simulations. The total SLS mass is $1\,\mathrm{M_\odot}$
in all cases. Stacked areas show the fraction of SLS mass bound to the SMBH
(blue), captured by the WD to form the WDDE (grey), and ejected from the
system on unbound trajectories (red). The vertical dashed lines mark the
boundaries of the collisional windows ($\varphi_0 \approx 26^\circ$--$32.5^\circ$
and $206^\circ$--$212.5^\circ$), within which the WD-captured fraction peaks and
the ejected fraction reaches both its minimum and maximum.}
    \label{fig:mass_budget}
\end{figure*}

Figure~\ref{fig:mass_budget} summarises the mass budget at $t = 53\,\mathrm{h}$
for all simulations as a function of $\varphi_0$, showing the SLS mass
partitioned into three stacked components: the fraction bound to the SMBH
(which will eventually fall back and power the TDE), the fraction captured by
the WD to form the WDDE, and the fraction ejected from the system on unbound
trajectories.

The overall trend follows the expectations from three-body dynamics: where the three-body calculations predict the SLS to be bound, the SMBH-bound fraction dominates, averaging $\sim 55\%$ over $\varphi_0 \in (40^\circ, 200^\circ)$ against $\sim 45\%$ ejected, and peaking at about $90\%$ near $\varphi_0 \simeq 31^\circ$. Symmetrically, where the three-body calculations predict the SLS to be unbound, the ejected fraction dominates, with the partition inverting over $\varphi_0 \in (220^\circ, 380^\circ)$, so that the two branches are close to, but not exactly, mirror images of one another. This agrees with the analytic expectation of Section~\ref{sec:ecc_TDE}: the SLS eccentricities span $e \simeq 0.97$--$1.03$ (Table~\ref{tab:sim_summary}), straddling the transition interval of equation~\eqref{eq:ecc_window}, and equation~\eqref{eq:bound_fraction} returns a mean bound fraction of $0.50$ across the suite.

The most striking departures from this general behaviour occur at the centres of the collisional windows ($\varphi_0 \simeq 28.4^\circ$ and $208.4^\circ$), where the ejected fraction reaches a minimum of only $\sim 5\%$ and the SMBH-bound fraction is likewise suppressed. This suppression coincides with the maximum of the WD-captured fraction, which peaks around the collisional windows as the direct WD--SLS collision at pericenter enables substantial mass transfer onto the WD.
In all such cases the closest WD--SLS approach occurs shortly after pericenter, by
$\Delta t \simeq 10^{2}$--$10^{3}$ s; near the centre of the window this is short compared with
the SLS dynamical time ($\simeq 1.6\times10^{3}$ s) and the WD meets an essentially intact star,
whereas towards the edges the two become comparable and it traverses already-expanding material.

Beyond the collisional windows the captured fraction declines but does not vanish at once:
mass transfer persists wherever the three-body minimum separation satisfies
$a_{\rm min} \lesssim 7\,R_\odot$, that is for $\varphi_0 \simeq 17^\circ$--$40^\circ$ and its
$180^\circ$ complement, and is identically zero outside. Capture therefore occupies $\sim 12\%$
of orientations, the collisional windows accounting for about a third of that; for the remaining
$\sim 88\%$ the WD emerges unchanged. The threshold is bracketed by S37 and S40
($\Delta M = 7\times10^{-2}$ and $10^{-3}\,M_\odot$), giving $\pm2.5^\circ$ on each edge, and the
boundaries may shift slightly with SPH resolution

\subsection{Formation and structure of the WDDE}
\label{subsec:wdde_formation}
\begin{figure*}
    \centering
    \includegraphics[width=\linewidth]{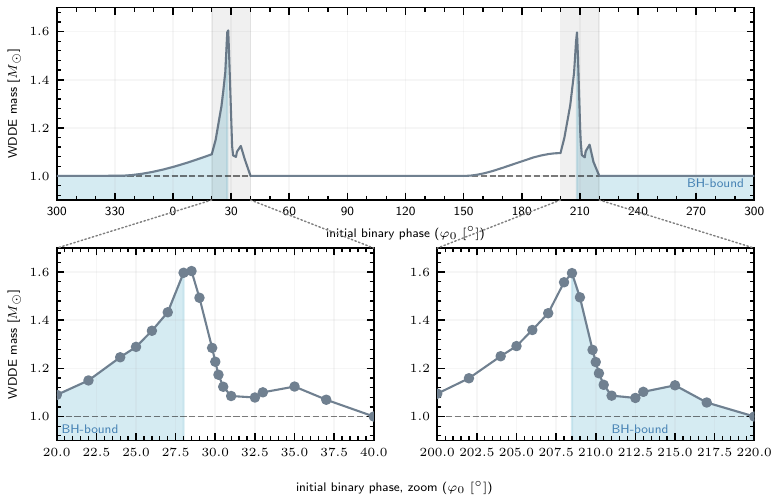}
    \caption{Total mass of the WDDE as a function of the initial binary phase $\varphi_0$, comprising the original WD ($1\,\mathrm{M_\odot}$, horizontal dashed line) plus the SLS material captured during the encounter. The blue shaded region marks configurations in which the WDDE remains bound to the SMBH.}
    \label{fig:wd_mass}
\end{figure*}

As discussed in Section~\ref{subsec:global_evolution}, the material captured by the WD during the encounter forms an extended, non-degenerate envelope that is still relaxing by the end of our simulations. The mass contained in this envelope varies systematically with $\varphi_0$, as shown in Figure~\ref{fig:wd_mass}, reaching its maximum at the centres of the collisional windows, where direct WD--SLS collisions enable the most efficient mass transfer. The resulting masses and radii of the WDDE for all cases are reported in the final two columns of Table~\ref{tab:sim_summary}.

Following the methodology detailed in Appendix~\ref{app:bound_mass_wd}, these values are calculated by identifying the gas particles that remain gravitationally bound to the WD core. Specifically, a particle is considered part of the WDDE if its relative orbital eccentricity $e_i < 1$. The characteristic radius is then defined as the distance enclosing 95$\%$ of this total bound mass ($R_{95}$), providing a robust measure of the system's spatial extent, for most encounters.

The formation of the WDDE is also reflected in the narrow blue peak visible in the eccentricity distributions of Figure~\ref{fig:histograms_all}, where a fraction of the SLS debris is concentrated within a small range of eccentricities centered on the WD orbit. This indicates that the captured debris is progressively relaxing into a gravitationally bound configuration around the WD.

In Figure~\ref{fig:wd_mass}, secondary peaks are also apparent at $\varphi_0 \approx 35^\circ$ and $215^\circ$, reaching values of up to $\sim 1.2\,M_\odot$. These peaks occur outside the collision window ($26^\circ \lesssim \varphi_0 \lesssim 32.5^\circ$), where the global maxima are found. Closer inspection indicates that they are numerical rather than physical, marking the limit of our binding criterion. At these phases the energy condition of Appendix~\ref{app:bound_mass_wd} does flag bound particles, but instead of relaxing into an approximately spherical envelope they remain organised as a narrow tidal bridge connecting the WD to the expanding debris stream. This material is still being stripped when our runs end, and is resolved by few SPH particles, a regime in which the method itself becomes unreliable. We therefore treat the WDDE masses at these two phases as upper limits rather than evidence for a distinct capture channel.

The structure of the WDDE at the end of our simulations can be broadly characterised as an unresolved degenerate core (the original WD) surrounded by an extended envelope of solar-like material. Where capture occurs at all, the WD retains between $\sim0.1\%$ and $\sim60\%$ of the SLS mass, the largest fractions coming from the near head-on collisions at the centre of the window.

\begin{figure*}
    \centering
    \includegraphics[width=\linewidth]{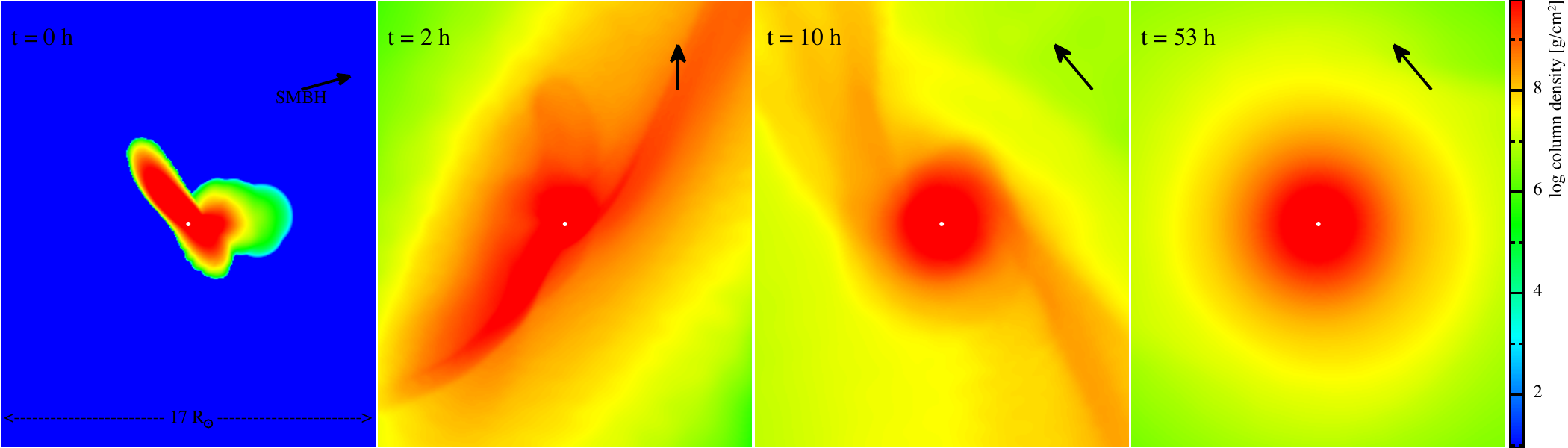}
    \caption{Column density maps for simulation S28.5 at $t = 0$, 2, 10, and 53 h. The black arrow marks the position of the SMBH, while the white point indicates the WD. Rendered using \texttt{SPLASH} \citep{2007SPLASH}.}
    \label{fig:collisonS28}
\end{figure*}

The temporal evolution of mass capture is illustrated in Figure~\ref{fig:collisonS28}, which presents column density maps for simulation S28.5, the case with the largest captured mass fraction. Mass capture occurs predominantly within the first few hours after pericenter passage, when the debris density is highest.

At $t = 0$\,h, the WD is fully embedded within the tidally disrupted SLS. By $t = 2$\,h, the system has exited the SLS tidal radius, and the densest debris has begun to concentrate around the WD. At $t = 10$\,h, although the SLS debris has expanded to scales of $\sim 600\,R_\odot$, most of the mass remains concentrated within $\sim 20\,R_\odot$, corresponding to material gravitationally bound to the WD. At this stage, a well-defined overdensity has formed around the WD, although two extended tidal tails are still present.
By $t = 53$\,h, the tidal tails have dissipated and the system has settled into a roughly spherical, extended envelope around the WD, with a characteristic radius of $\sim 5.5\,R_\odot$ and a total mass of $\sim 1.6\,M_\odot$.

A structurally analogous object arises in a different context: a red giant is itself a degenerate core surrounded by a loosely bound envelope, and simulations of partial tidal disruptions of giants find that the core, together with a fraction of the envelope, can survive the encounter and in some cases escape to infinity \citep{navarro2024giant}. The WDDE is assembled by the opposite route, acquiring an envelope rather than retaining its own, but raises closely related questions about the evolution of a degenerate core embedded in an extended envelope.

The captured mass fractions we find are broadly consistent with those reported
by \citet{van20243d} for isolated WD--main-sequence collisions, who
find that head-on encounters can transfer between $45\%$ and $65\%$ of the
companion mass onto the WD. Our results span a wider range of mass transfer, which we attribute
to two factors: the presence of the SMBH, which tidally disrupts the SLS and
limits the fraction of its mass available for capture, and the variety of
impact geometries in our sample, from nearly head-on collisions at the centre
of the collisional window to off-centre impacts at its boundaries where the WD
grazes the outer layers of the SLS rather than its core. In the off-centre
regime, \citet{van20253d} report capture fractions of $50\%$--$80\%$
for star-WD collisions, substantially higher than our values of $\sim 0.1\%$--$60\%$
for the same geometry; this difference is consistent with the SMBH tidal
disruption reducing the available mass reservoir in our simulations.

\subsection{Fate of the WDDE}
\label{subsec:wdde_fate}

While our hydrodynamical simulations provide a detailed look at the initial capture process, our treatment of the WD as a point particle neglects possible nuclear reactions and detailed thermodynamic dissipation. Consequently, our assessment of the long-term evolution of the WDDE is based on the total bound mass and characteristic envelope size rather than a comprehensive physical model. Within these numerical limitations, several evolutionary scenarios remain noteworthy.

Specifically, for encounters within the collisional window, the

total mass of the WDDE frequently exceeds $1.4\,M_\odot$. We stress that this
does not imply that the degenerate core exceeds the Chandrasekhar mass. As described
above, the captured material settles into an extended, non-degenerate envelope around
a core that retains its original mass, so the relevant question for the fate of the
system is not the total mass but how much of that envelope is eventually incorporated
into the core. This is set by the competition between accretion from the envelope and
mass loss, and cannot be resolved by our simulations, which end long before the envelope
has had time to evolve thermally.
We do not expect an immediate Type~Ia supernova explosion, as the mass transfer occurs
over a few hours, characterizing a highly dynamical, non-equilibrium accretion process.
It is more probable that once the accretion rate

onto the core exceeds the rate at which the accreted material can be burned stably,
which for a WD of this mass is of order $10^{-6}\,M_\odot\,\mathrm{yr^{-1}}$,
a fraction of the material is expelled through nova-like episodes.
Depending on the final mass retention, the WDDE might eventually undergo a thermonuclear explosion or settle into a peculiar red-giant--like configuration, as suggested by previous literature \citep[e.g.,][]{shara1977white,shara1986hydrodynamic, van20243d}. 

Finally, it is important to emphasize that the envelope has not reached hydrostatic equilibrium by the end of our simulations ($t = 53\,\mathrm{h}$) and will continue to relax on much longer timescales. A definitive assessment of its ultimate fate thus requires dedicated long-term modelling incorporating additional physics, which we defer to future work.

\subsection{Future pericenter passages and repeating TDEs}
\label{subsec:repeating}
\begin{table*}
\centering
\begin{tabular}{l c c c l c c c}
\hline
\multicolumn{3}{c}{\textbf{Unbound WD/WDDE}} & \vrule & \multicolumn{4}{c}{\textbf{Bound WD/WDDE}} \\
\hline
Simulation & $v_{\infty}$ & Mass & \vrule & Simulation & Orbital Period & Mass & Impact parameter \\
 & [km s$^{-1}$] & [$M_\odot$] & \vrule & & [yr] & [$M_\odot$] & \\
\hline
S28.5 & 1491.85 & 1.6047 & \vrule & S20 & 1.642 & 1.09 & 34.33 \\
S29 & 3420.81 & 1.4937 & \vrule & S22 & 1.348 & 1.1501 & 34.65 \\
S29.8 & 5249.15 & 1.2854 & \vrule & S24 & 1.176 & 1.2463 & 28.59 \\
S30 & 5561.80 & 1.2278 & \vrule & S25 & 1.038 & 1.2896 & 25.38 \\
S30.2 & 5734.55 & 1.1738 & \vrule & S26 & 0.957 & 1.3565 & 15.26 \\
S30.5 & 5709.78 & 1.1239 & \vrule & S27 & 1.054 & 1.4334 & 10.71 \\
S31 & 5410.09 & 1.0854 & \vrule & S28 & 8.692 & 1.5976 & 8.82 \\
S32.5 & 4113.37 & 1.0792 & \vrule & S208.5 & 5.968 & 1.5961 & 5.48 \\
S33 & 3793.18 & 1.1006 & \vrule & S209 & 0.689 & 1.4955 & 5.85 \\
S35 & 3150.27 & 1.1247 & \vrule & S209.8 & 0.177 & 1.2771 & 5.03 \\
S37 & 2873.18 & 1.0699 & \vrule & S210 & 0.153 & 1.2266 & 5.20 \\
S40 & 2290.48 & 1.0001 & \vrule & S210.2 & 0.140 & 1.1798 & 4.58 \\
S70 & 1576.39 & 1.0 & \vrule & S210.5 & 0.140 & 1.1320 & 4.85 \\
S90 & 1648.10 & 1.0 & \vrule & S211 & 0.164 & 1.0869 & 4.80 \\
S100 & 1697.40 & 1.0 & \vrule & S212.5 & 0.366 & 1.0774 & 10.99 \\
S120 & 1800.84 & 1.0 & \vrule & S213 & 0.478 & 1.1027$^{\star}$ & 12.1947$^{\star}$ \\
S150 & 1939.00 & 1.0 & \vrule & S215 & 0.817 & 1.1298$^{\star}$ & 12.2323$^{\star}$ \\
S200 & 2492.08 & 1.095 & \vrule & S217 & 1.093 & 1.0584$^{\star}$ & 12.2106$^{\star}$ \\
S202 & 2669.43 & 1.1591 & \vrule & S220 & 2.106 & 1.0001 & 0.0122 \\
S204 & 2785.77 & 1.2507 & \vrule & S250 & 6.035 & 1.0 & 0.0123 \\
S205 & 2920.20 & 1.2928 & \vrule & S270 & 5.330 & 1.0 & 0.0125 \\
S206 & 2994.72 & 1.3597 & \vrule & S280 & 4.902 & 1.0 & 0.0126 \\
S207 & 2901.04 & 1.4295 & \vrule & S300 & 4.1605 & 1.0 & 0.0127 \\
S208 & 1696.25 & 1.5581 & \vrule & S330 & 3.367 & 1.0 & 0.0127 \\
\hline
\end{tabular}
\caption{Summary of the WD/WDDE outcomes. Left: unbound cases with asymptotic velocities $v_{\infty}$ and masses. Right: cases that remain gravitationally bound to the SMBH, with their orbital periods, masses and resulting impact parameters. Values marked $^{\star}$ are upper limits; see the note to Table~\ref{tab:sim_summary}.}
\label{tab:wdde_combined_horizontal}
\end{table*}


The different outcomes of the WDDE in our simulations are summarized in Table~\ref{tab:wdde_combined_horizontal}, distinguishing between cases in which the object remains gravitationally bound to the SMBH and those in which it is ejected as a hypervelocity object. For bound cases the new pericenter stays within $\sim80-84\,R_\odot$ of the original, but the new impact parameters span a much wider range ( $\beta_\mathrm{WDDE}\simeq4.6$--34.6),
from S210.2 at the lower end to S22 at the upper end.

Since the WDDE is a hybrid object comprising a WD core and an extended envelope, the SMBH tidal field (notwithstanding the large $\beta$ values) is expected to strip a fraction of the envelope at each subsequent pericenter passage. This process likely leaves behind a dense core and could give rise to recurrent accretion episodes. The recurrence timescale is governed by the orbital period of the WDDE, which depends on the orbital energy imparted during the encounter and thus varies systematically with $\varphi_0$. For the parameters explored in our simulations, this period ranges from approximately 50 days in the most tightly bound cases, such as S210.2 and S210.5, to 6 and 9 years for encounters S208.5 and S28, respectively, which sit right at the transition between the bound and unbound regimes. This applies only to a subset of encounters: the WD/WDDE remains bound for half of all orientations, and of those only the ones adjacent to the collisional windows retain an envelope. For the rest the companion survives as a bare WD of $1\,M_\odot$ with $\beta_\mathrm{WD} \simeq 0.01$, leaving nothing for the SMBH tidal field to remove.

Where an envelope is present, the progressive stripping of the WDDE at each pericenter passage would produce repetitive signals similar to those of a TDE, but distinct from the initial disruption in both spectral properties and recurrence pattern. Such behaviour resembles repeating partial TDEs, where a bound stellar remnant undergoes multiple passages and generates flares separated by the orbital period \citep[e.g.,][]{pasham2024potential,somalwar2025first}. If multiple stripping episodes occur, a sequence of flares with decreasing amplitude is expected as the envelope is gradually depleted \citep[e.g.,][]{broggi2024repeating,chen2024fate}.

\subsection{Unbound hypervelocity WD/WDDE}
\label{subsec:hypervelocity}

In contrast, in simulations where the WD/WDDE acquires sufficient energy to escape the SMBH (see Table~\ref{tab:wdde_combined_horizontal}), the system is ejected with asymptotic velocities of several thousand km s$^{-1}$. These objects correspond to hyperbolic objects produced by energy exchange during the encounter.

The resulting velocities are comparable to those observed in hypervelocity stars \citep[e.g.,][]{brown2015hypervelocity}, suggesting that this mechanism may contribute to their population. Unlike the classical Hills mechanism, in which a star is ejected following the tidal splitting of a binary without significant hydrodynamical interaction, the ejected object in this scenario is not always an intact star: for phases within the capture range it carries a debris envelope acquired during the encounter and escapes as a WDDE, with potentially distinct observational signatures. This applies to roughly one ejection in eight; for the remainder the companion escapes as a bare WD, indistinguishable from the classical Hills outcome. Observationally, hypervelocity stars with evolved stellar properties, including red giant candidates, have been identified \citep[e.g.,][]{hattori2025discovery}, highlighting the diversity of objects that may be produced in such extreme dynamical interactions.

In the former cases, these objects may evolve into peculiar configurations, potentially observable as objects with inflated envelopes or as transients associated with the expansion and cooling of the bound material. The detection of such systems would provide indirect evidence of binary--SMBH encounters.

A detailed assessment of the long-term evolution and observational signatures of these objects is beyond the scope of this work and will be explored in future studies.

\section{A menagerie of Tidal Disruption Events}
\label{sec:menagerie}

The fallback rate of stellar debris onto the SMBH is the primary diagnostic available from our
simulations, encoding directly information about the orbital energy distribution of the disrupted material. While the classical single-star parabolic case is characterized by a rapid rise to peak followed by the canonical $t^{-5/3}$ power-law decay, the binary encounters studied here exhibit significant departures from this baseline. These deviations depend primarily on two key aspects of the interaction: whether the WD and SLS undergo a collision near pericenter, and whether the WD/WDDE remains bound to the SMBH or is ejected following tidal separation.

\begin{figure}
    \centering
    \begin{overpic}[width=\linewidth]{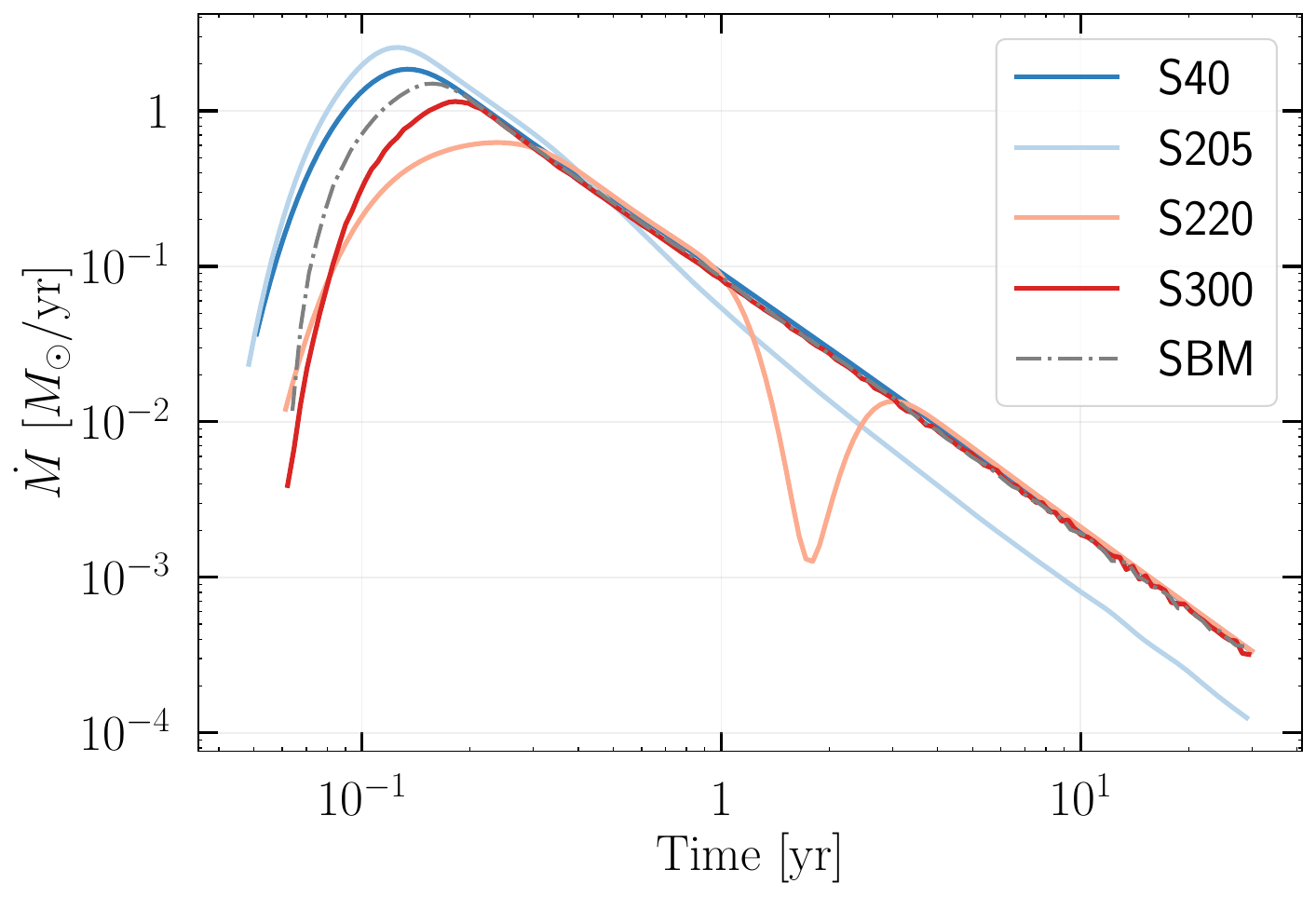}
        \put(100,150){\color{blue}\textbf{eTDE}}
        \put(80,105){\color{red}\textbf{hTDE}}
    \end{overpic}
    \caption{Fallback rate curves for representative eTDE (blue) and hTDE (red) cases. The eTDE cases S40 and S205 follow the expected behaviour for eccentric disruptions, while S300, which lies outside the capture range, follows the theoretical hyperbolic profile. In contrast, S220, whose phase falls within the capture range, exhibits different fallback-rate behaviour compared to the other cases. The gray dashed line corresponds to the benchmark simulation.}
    \label{fig:etde_htde_cases}
\end{figure}

The resulting diversity of fallback rate behaviours can be naturally organised into the two primary classes introduced in Section~\ref{sec:ecc_TDE}: elliptical TDEs (eTDEs) and hyperbolic TDEs (hTDEs), as illustrated in Figure~\ref{fig:etde_htde_cases}. The former corresponds to encounters in which the WD/WDDE acquires sufficient energy to escape the SMBH. As established by energy conservation during binary separation, the stellar debris is predominantly placed on tightly bound elliptical orbits around the SMBH, leading to earlier return times and enhanced peak fallback rates compared to the parabolic baseline.

Conversely, hTDEs occur when the WD/WDDE remains gravitationally bound to the SMBH. In this configuration, the SLS debris receives a positive energy kick, placing it on less-bound or marginally unbound orbits. This produces a fallback rate characterized by a delayed onset and lower peak fallback rates than the SBM case. Over most of this range the bound companion survives as a bare WD, with no envelope to strip; only for phases within the capture range does it become a WDDE (Table~\ref{tab:wdde_combined_horizontal}). Repeating TDE (rTDE) signatures are therefore expected only in that subset (Section~\ref{subsec:repeating}), arising from the same encounter as the initial fallback but as a physically distinct, later phenomenon.

Because the energy exchanged during the separation is antisymmetric between the two components, the SLS kick changes sign at a single phase, which we locate at $\varphi_c \simeq 28.4^\circ$ by interpolating the debris eccentricity of Table~\ref{tab:sim_summary} through unity. Its $180^\circ$ complement is measured independently at $208.4^\circ$, the two crossings being separated by $179.94^\circ$. These two points divide the phase space into exact halves according to whether the WD/WDDE remains bound to the SMBH. The fallback classes defined below do not coincide with this split, since the collisional windows straddle both crossings: eTDEs and hTDEs each account for
$48\%$ of all binary orientations, and collision-driven TDEs (cTDEs), in which the WD and the SLS physically collide near pericenter, for the remaining $4\%$. We note that $\varphi_c$ is interpolated between simulations separated by $0.5^\circ$ and may shift slightly with resolution.
Within each class, however, additional structure emerges depending on whether the encounter is collisional or non-collisional; these subcategories will be explored in more detail in the following sections.
Quantitative values of the peak time, peak fallback rate and late-time decay index for every run are collected in Table~\ref{tab:fallback_summary}.

To estimate the peak luminosity, we assume it scales proportionally with the fallback rate, $L_{\rm peak} \sim \eta \dot{M}_{\rm peak} c^2$, adopting a fiducial radiative efficiency of $\eta = 0.1$. This estimate should be interpreted with caution, as $\eta$ encapsulates significant uncertainties in the emission mechanisms of TDEs. Therefore, the derived values provide only order-of-magnitude estimates of the peak luminosity \citep[e.g.,][]{coughlin2023luminosity}.
\begin{table*}
\centering
\small
\caption{Fallback properties for the binary--SMBH encounter simulations. Columns show: simulation identifier, peak fallback rate time $t_{\rm peak}$, peak fallback rate $\dot{M}_{\rm peak}$, peak luminosity $L_{\rm peak}$ (assuming $\eta=0.1$), power-law slope $\alpha$, and TDE classification. The slope $\alpha$ is obtained by fitting a power law to the late-time fallback curve, in the interval from 1 to 40 years. The reference simulation SBM is repeated at the top of the second block for visual reference.}
\label{tab:fallback_summary}
\setlength{\tabcolsep}{5pt}
\renewcommand{\arraystretch}{1.05}
\begin{tabular}{lccccc | lccccc}
\hline
\hline
Simulation & $t_{\rm peak}$ & $\dot{M}_{\rm peak}$ & $L_{\rm peak}$ & $\alpha$ & TDE type & Simulation & $t_{\rm peak}$ & $\dot{M}_{\rm peak}$ & $L_{\rm peak}$ & $\alpha$ & TDE type \\
 & [yr] & [$M_\odot\,{\rm yr}^{-1}$] & [$10^{45}$ erg s$^{-1}$] & & & & [yr] & [$M_\odot\,{\rm yr}^{-1}$] & [$10^{45}$ erg s$^{-1}$] & & \\
\hline
\textbf{SBM} & \textbf{0.158} & \textbf{1.50} & \textbf{8.54} & \textbf{1.662} & \textbf{Classic} & \textbf{SBM} & \textbf{0.158} & \textbf{1.50} & \textbf{8.54} & \textbf{1.662} & \textbf{Classic} \\
S20 & 0.214 & 0.626 & 3.57 & 1.658 & hTDE & S200 & 0.128 & 2.01 & 11.5 & 1.674 & eTDE \\
S22 & 0.209 & 0.467 & 2.65 & 1.600 & hTDE & S202 & 0.129 & 2.090 & 11.84 & 1.740 & eTDE \\
S24 & 0.227 & 0.347 & 1.97 & 1.625 & hTDE & S204 & 0.128 & 2.318 & 13.13 & 1.732 & eTDE \\
S25 & 0.231 & 0.260 & 1.48 & 1.660 & hTDE & S205 & 0.127 & 2.56 & 14.6 & 1.731 & eTDE \\
S26 & 0.254 & 0.170 & 0.97 & 1.600 & cTDE & S206 & 0.121 & 3.21 & 18.3 & 1.715 & cTDE \\
S27 & 0.312 & 0.0737 & 0.42 & 1.601 & cTDE & S207 & 0.108 & 3.29 & 18.7 & 1.671 & cTDE \\
S28 & 0.047 & 0.319 & 1.82 & 1.517 & cTDE & S208 & 0.106 & 1.20 & 6.85 & 1.996 & cTDE \\
S28.5 & 0.0426 & 1.41 & 8.05 & 1.985 & cTDE & S208.5 & 0.117 & 0.936 & 5.34 & 1.592 & cTDE \\
S29 & 0.0389 & 5.12 & 29.2 & 1.704 & cTDE & S209 & 0.121 & 0.544 & 3.10 & 1.632 & cTDE \\
S29.8 & 0.0492 & 16.2 & 92.3 & 1.677 & cTDE & S209.8 & 0.0359 & 0.305 & 1.74 & 1.609 & cTDE \\
S30 & 0.0567 & 22.6 & 129 & 1.717 & cTDE & S210 & 0.0343 & 0.265 & 1.51 & 1.648 & cTDE \\
S30.2 & 0.0667 & 31.3 & 179 & 1.726 & cTDE & S210.2 & 0.0366 & 0.213 & 1.22 & 1.649 & cTDE \\
S30.5 & 0.0832 & 31.0 & 177 & 1.669 & cTDE & S210.5 & 0.0419 & 0.151 & 0.861 & 1.687 & cTDE \\
S31 & 0.127 & 19.4 & 111 & 1.695 & cTDE & S211 & 0.0511 & 0.0869 & 0.496 & 1.625 & cTDE \\
S32.5 & 0.110 & 3.31 & 18.9 & 1.848 & cTDE & S212.5 & 0.143 & 0.0312 & 0.178 & 1.544 & cTDE \\
S33 & 0.114 & 2.83 & 16.1 & 1.718 & eTDE & S213 & 0.166 & 0.0309 & 0.176 & 1.367 & hTDE \\
S35 & 0.122 & 2.222 & 12.59 & 1.654 & eTDE & S215 & 0.182 & 0.101 & 0.58 & 1.697 & hTDE \\
S37 & 0.130 & 2.012 & 11.40 & 1.659 & eTDE & S217 & 0.196 & 0.267 & 1.51 & 1.700 & hTDE \\
S40 & 0.134 & 1.85 & 10.6 & 1.662 & eTDE & S220 & 0.235 & 0.625 & 3.56 & 1.671 & hTDE \\
S70 & 0.142 & 1.838 & 10.41 & 1.670 & eTDE & S250 & 0.181 & 1.214 & 6.88 & 1.687 & hTDE \\
S90 & 0.142 & 1.870 & 10.59 & 1.657 & eTDE & S270 & 0.181 & 1.201 & 6.80 & 1.649 & hTDE \\
S100 & 0.141 & 1.893 & 10.72 & 1.671 & eTDE & S280 & 0.181 & 1.180 & 6.69 & 1.655 & hTDE \\
S120 & 0.137 & 1.945 & 11.02 & 1.672 & eTDE & S300 & 0.182 & 1.149 & 6.51 & 1.668 & hTDE \\
S150 & 0.139 & 2.016 & 11.42 & 1.673 & eTDE & S330 & 0.186 & 1.096 & 6.21 & 1.657 & hTDE \\
\hline
\hline
\end{tabular}
\end{table*}

\subsection{Elliptical TDEs (eTDEs)}
\label{subsec:E-TDE}

These encounters occur for initial phases $\varphi_0\in(32.5^\circ, 206^\circ)$ comprising $48\%$ of all binary orientations. In this regime, the WD/WDDE is ejected from the system, while the stellar debris from the SLS is placed on tightly bound orbits around the SMBH. As a result, elliptical TDEs exhibit systematic deviations from the classical parabolic case. In particular, they are characterised by an earlier onset of fallback rate, significantly higher peak fallback rates, and an overall increase in the total accreted mass. These features reflect the more negative energy distribution of the debris compared to the SBM scenario (see Figure~\ref{fig:histograms_all}).

These three signatures are clearly visible in Figure~\ref{fig:etde_htde_cases} for the representative cases S40 S205 (in blue), which provide the cleanest realisations of the eTDE class, in good agreement with theoretical predictions for eccentric disruptions (Section~\ref{sec:ecc_TDE}). These results demonstrate that tidal separation of stellar binaries provides a natural mechanism for injecting stars onto the bound orbits required for this class of event.

 The $\sim 12\%$ of eTDEs whose phases fall adjacent to the collisional regions (Section~\ref{subsec:mass_budget}) (e.g. S205, shown in the Figure~\ref{fig:etde_htde_cases}) exhibit a higher peak fallback rate and a more abrupt decline than the other eTDE cases. At later times, however, the fallback rate evolution converges towards the canonical power-law decay expected for parabolic TDEs.

It is worth noting that, for the vast majority of eTDEs, the orbital eccentricity of the disrupted star deviates only modestly from unity ($0.990 \lesssim e \lesssim 0.999$, Table~\ref{tab:sim_summary}), so the resulting fallback rate curves remain close to the standard parabolic case. A clean eTDE could therefore plausibly be confused with a standard parabolic TDE from a more massive progenitor or a deeper encounter \citep{guillochon2009three}; distinguishing these scenarios would require either a precise measurement of the fallback rate shape or independent constraints on the stellar population in the galactic nucleus.

\subsection{Hyperbolic TDEs (hTDEs)}
\label{subsec:RH-TDE}

These encounters correspond to the complementary range of the parameter space, occurring for $\varphi_0\in(0^\circ, 32.5^\circ)$ and $\varphi_0\in(206^\circ, 360^\circ)$, and accounting for $48\%$ of all binary orientations. In this regime, the WD/WDDE remains bound to the SMBH. By energy conservation during the binary separation, the SLS receives a positive energy kick, placing its debris on less bound or marginally unbound orbits compared to the parabolic case. As a result, the initial disruption exhibits the characteristic properties of a hyperbolic TDE (hTDE), including a delayed onset, a lower peak fallback rate, and a reduced total accreted mass relative to the SBM case.

The fallback rate curves for representative cases S300 and S220, shown with red lines in Figure~\ref{fig:etde_htde_cases}, broadly follow theoretical expectations for hyperbolic disruptions. S300 lies far from the collisional windows, where no mass is captured and the fallback rate follows the expected hyperbolic profile without further structure; this is the case for $\sim 88\%$ of hTDEs. In contrast, S220 falls within the capture range (e.g. S20, S25, S213, S215 and 220) and exhibits a distinctive additional feature in its fallback rate curve: a well-defined gap followed by a recovery. This gap arises from the presence of the bound WDDE, which travels alongside the returning debris stream and captures a significant fraction of the bound material before it can accrete onto the SMBH (Section~\ref{subsec:mass_budget}). The feature is therefore specific to hTDEs: in eTDEs the WD/WDDE is ejected and does not remain to intercept the returning stream, so capture leaves at most a minor imprint on the fallback curve.

A qualitatively similar feature has been reported in recent hydrodynamical simulations of binary disruptions involving two solar-type stars \citep{yu2025binary}, where a comparable dip in the fallback rate (see their Figure~11, Run 2) is instead attributed to a surviving merger remnant. In our case, however, this feature is naturally explained by the formation of a WDDE. Where a WDDE forms and remains bound, repeating flares are expected (Section~\ref{subsec:repeating}); where the companion stays a bare WD, no envelope is available to strip and no such activity is anticipated. In either case, however, the bound WD/WDDE continues to orbit through the debris and may perturb the accretion disc that forms later, offering a further route to quasi-periodic eruptions (QPEs)\citep{arcodia2024cosmic,suzuguchi2026quasi}.

\subsection{Collision-driven TDEs (cTDEs)}
\label{subsec:cTDE}

\begin{figure}
    \centering
    \includegraphics[width=\linewidth]{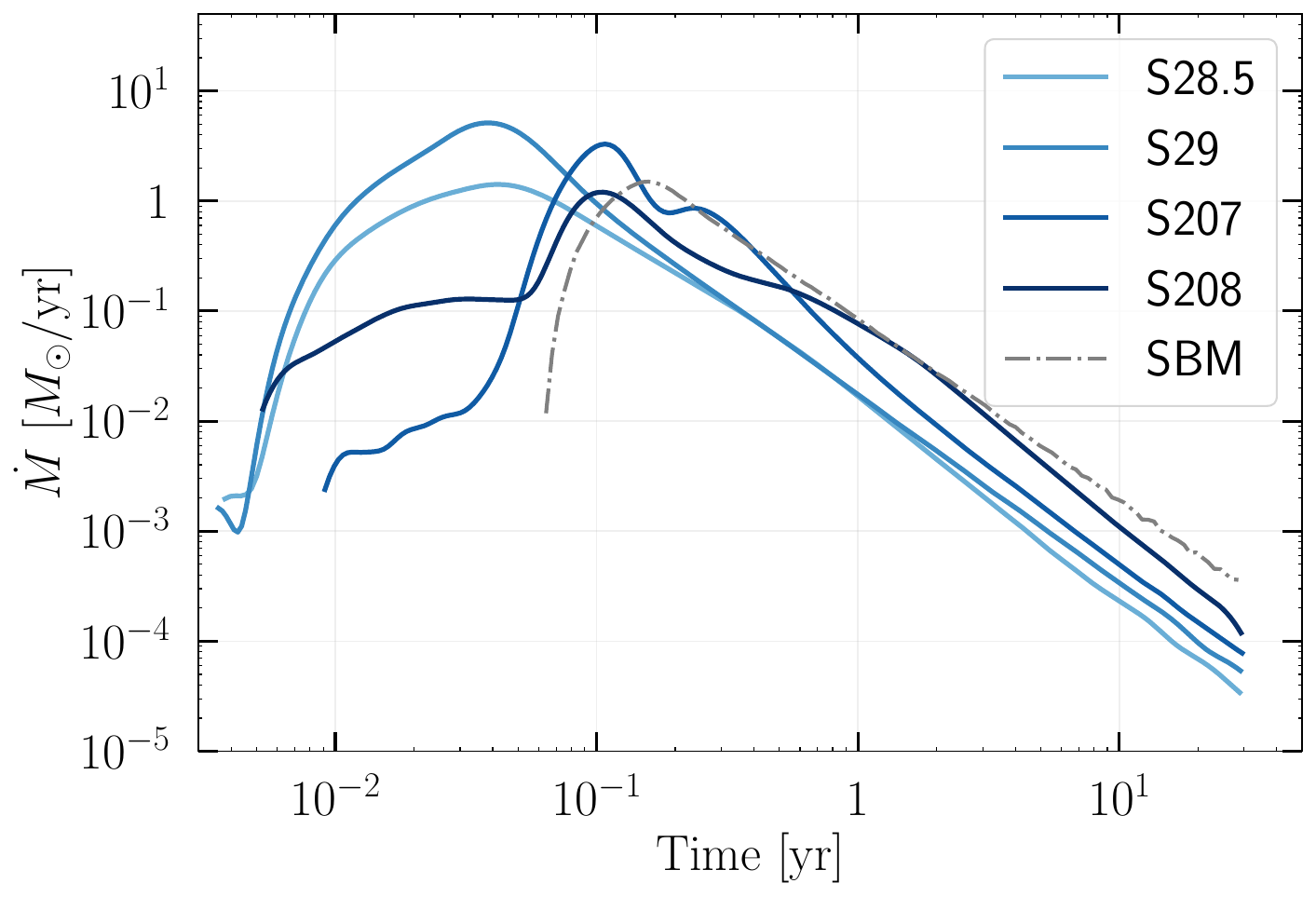}\\
     \includegraphics[width=\linewidth]{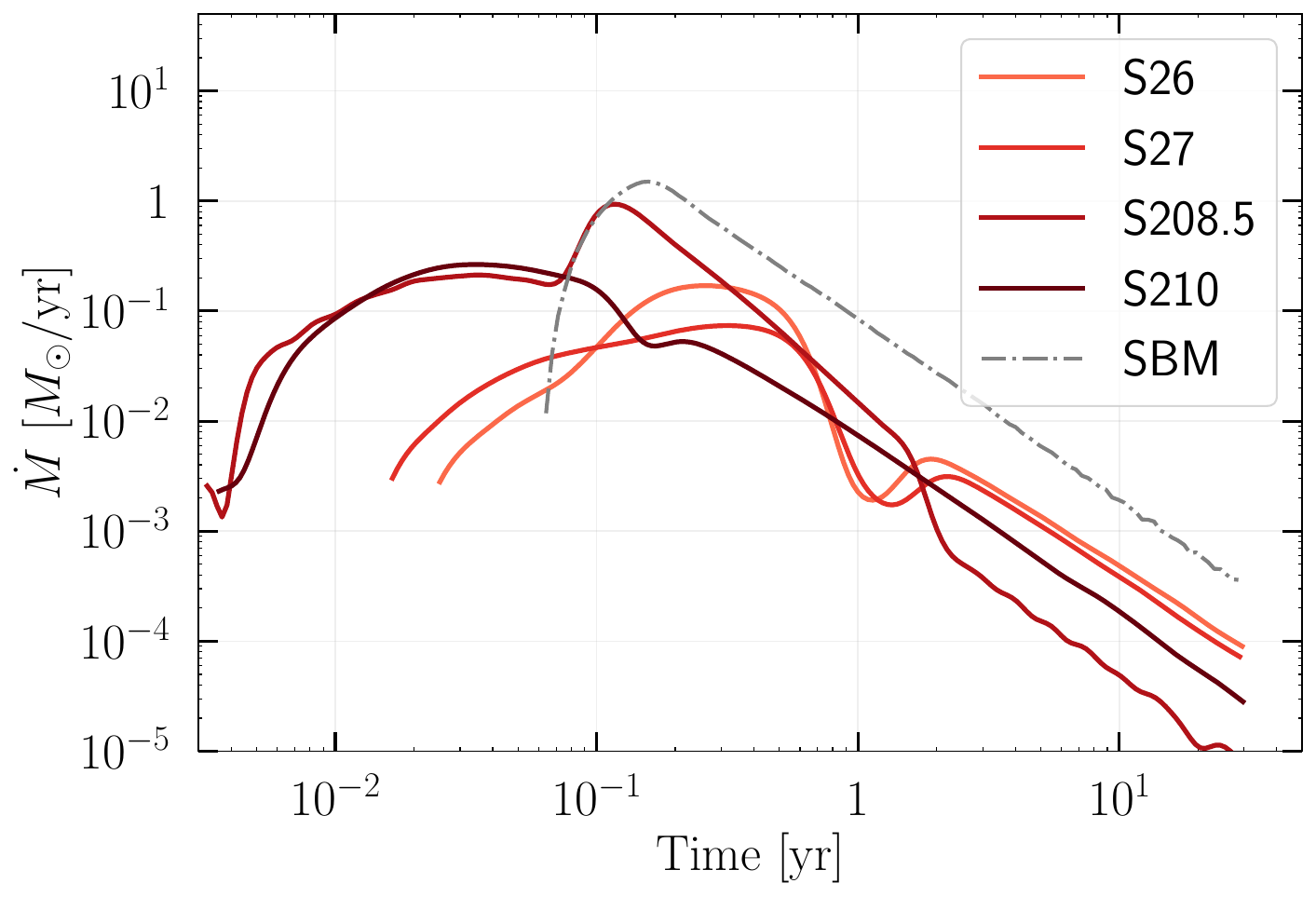}\\
    \caption{Fallback rate curve for representative collision-driven TDEs (cTDEs). The upper panel shows elliptical cTDEs (blue), including S28.5, S29, S207, and S208, while the lower panel shows hyperbolic cTDEs (red), including S26, S27, S208.5, and S210. In both panels, the gray dashed line shows the fallback rate of the single-star parabolic disruption (SBM) for reference. The cTDEs exhibit substantially more structured fallback-rate evolution than the canonical parabolic case.}
    \label{fig:collision_driven}
\end{figure}

Collision-driven TDEs (cTDEs) correspond to encounters in which a direct WD--SLS collision occurs near pericenter, producing a substantial redistribution of orbital energy within the debris stream. This interaction broadens the debris eccentricity distribution relative to the standard non-collisional cases and enhances the fraction of material placed onto more tightly bound orbits (see Figure~\ref{fig:histograms_all}). As a consequence, the fallback rate evolution is strongly modified, leading to earlier debris return times and significantly more structured fallback rate curves than those found in standard eTDE and hTDE events, as illustrated in the Figure~\ref{fig:collision_driven}.

These encounters occur within the collisional window, approximately spanning $\varphi_0\in(26^\circ, 32.5^\circ)$ and $\varphi_0\in(206^\circ, 212.5^\circ)$, and comprise $\sim 4\%$ of all orientations. In all cTDEs, the collision-induced redistribution of orbital energy generates noticeable deviations from the canonical single-star fallback rate evolution. In particular, the fallback rate can peak on timescales as short as $\sim0.034\,\mathrm{yr}$ after pericenter passage (S210 in Table~\ref{tab:fallback_summary}), substantially earlier than the $\sim0.16\,\mathrm{yr}$ characteristic of the SBM case.

We stress that cTDEs are not a third energetic class. Both sign changes of the SLS energy kick, at $\varphi_c$ and $\varphi_c +180^\circ$, lie inside their respective collisional windows, so each window is itself divided between the two classes; our collisional runs accordingly split evenly between bound and unbound debris, eight of each. What distinguishes cTDEs is the morphology of the fallback curve rather than the fate of the debris.
The fallback rate curves of cTDEs exhibit a wide variety of structured features, including sharper peaks, modified late-time slopes, and complex early-time variability. These signatures arise from the continued dynamical interaction between the debris and the surviving WDDE, which can either remain bound to or escape from the SMBH after the encounter.

Moreover, the collision-driven phenomenology evolves continuously across the phase space of the collisional window, producing a smooth transition between the behaviours associated with the elliptic and hyperbolic regimes. One consequence is that the timing diagnostic which separates clean events is lost: clean eTDEs and hTDEs occupy disjoint ranges of peak time ($t_{\rm peak} = 0.11$--$0.14$ yr and $0.17$--$0.24$ yr respectively), whereas both subsets of cTDEs begin near $0.034$ yr and overlap throughout. The ordering in peak rate nevertheless survives, bound debris still giving the higher values. Several representative cases (e.g. S207 and S210 in Figure~\ref{fig:collision_driven}) display pronounced departures from the smooth fallback rate evolution expected for the SBM case, providing a potential observational signature of their binary origin.

Table~\ref{tab:outcome_summary} summarizes the three mutually exclusive outcome classes identified in this work, along with their quantitative properties. Each row corresponds to a distinct range of initial binary phase $\varphi_0$ (assumed uniformly distributed over $360^\circ$), so the listed fractions sum to unity. The eTDE and hTDE classes together account for $\sim 92\%$ of the phase space and produce clean fallback curves, distinguished mainly by the timing and height of the peak. Within the hTDE population, a narrow sliver of phase ($\sim 4\%$) instead produces a gap in the fallback curve before the light curve resumes the delayed, suppressed hTDE morphology. The remaining $\sim 4\%$ of phase space, confined to narrow windows, gives rise to cTDEs, for which the fallback rate timing is no longer a reliable diagnostic. We note that our simulation suite deliberately oversamples these collisional windows, so the sampling of each class in our library is not proportional to its share of phase space.

\begin{table*}
\centering

\caption{Summary of the three mutually exclusive outcome classes of binary--SMBH
encounters identified in this work, together with the classic parabolic TDE
benchmark. Fractions refer to the $360^\circ$ of initial binary phase $\varphi_0$,
assumed uniformly distributed, and sum to $100\%$ across the three binary-disruption
classes; the benchmark row is not part of this partition. Bold values of $t_{\rm peak}$
and $\dot{M}_{\rm peak}$ are medians over each class, with the full range given
below; $\alpha$ is the late-time decay index. All values are taken from
Table~\ref{tab:fallback_summary}.}
\label{tab:outcome_summary}
\begin{tabular}{lcccccm{2.65cm}}
\hline
Class & Morphology & Fraction & $t_{\rm peak}$ [yr] & $\dot{M}_{\rm peak}$ [$M_\odot\,{\rm yr}^{-1}$] & $\alpha$ & Characteristic signature \\
\hline
TDE & -- & -- &
  $\mathbf{0.158}$ &
  $\mathbf{1.50}$ &
  $\mathbf{1.662}$ & Classic parabolic case \\
\hline
eTDE & Clean & $48\%$ &
  \valrango{$\mathbf{0.133}$}{0.114--0.142} &
  \valrango{$\mathbf{2.01}$}{1.83--2.83} &
  \valrango{$\mathbf{1.672}$}{1.654--1.74} & Earlier onset,\newline enhanced peak\\
\hline
\multirow[c]{2}{*}[-2pt]{hTDE} & Clean & $44\%$ &
  \valrango{$\mathbf{0.184}$}{0.181--0.186} &
  \valrango{$\mathbf{1.180}$}{1.096--1.214} &
  \valrango{$\mathbf{1.657}$}{1.649--1.687} & Delayed onset,\newline suppressed peak \\[10pt]
 & Gap & $\sim 4\%$ &
  \valrango{$\mathbf{0.201}$}{0.166--0.235} &
  \valrango{$\mathbf{0.307}$}{0.101--0.626} &
  \valrango{$\mathbf{1.659}$}{1.367--1.7} & Delayed onset,\newline suppressed peak, gap \\
\hline
cTDE & Collisional & $\sim 4\%$ &
  \valrango{$\mathbf{0.075}$}{0.034--0.312} &
  \valrango{$\mathbf{1.07}$}{0.03--31.3} &
  \valrango{$\mathbf{1.67}$}{1.52--1.99} & Collision-driven,\newline complex structure \\
\hline
\end{tabular}
\end{table*}

 Every cTDE forms a WDDE, and in those cases where it remains bound the envelope may be stripped further at subsequent pericenter passages, powering the delayed or recurrent activity discussed in Section~\ref{subsec:repeating}.

Furthermore, the peak fallback rates found in some of these cases exceed the single-star benchmark by more than an order of magnitude (Table~\ref{tab:fallback_summary}). If a modest fraction of this material were accreted promptly, the resulting flow would be strongly super-Eddington. This inference depends on the uncertain fallback-to-accretion mapping discussed in Section~\ref{sec:theory}, however, so the tabulated rates are best regarded as upper limits on $\dot{M}_{\rm acc}$. TDEs that significantly exceed the Eddington accretion rate are thought to be capable of launching relativistic jets, as observed in several jetted TDE candidates such as Swift~J1644+57 \citep{burrows2011relativistic}. Subject to the caveat above, some of our events reach inferred luminosities of up to $\sim 10^{46}\,\mathrm{erg\,s^{-1}}$, which would place them in a regime where jet production may be possible. Having identified these candidate cases, a more detailed investigation of jet formation, including the role of magnetic fields and black hole spin, can be carried out in future \citep[e.g., ][]{DECOLLE2020101538}.
\subsubsection{Surviving SLS cores}
\label{subsec:surviving_sls_cores}

\begin{figure}
    \centering
    \includegraphics[width=\linewidth]{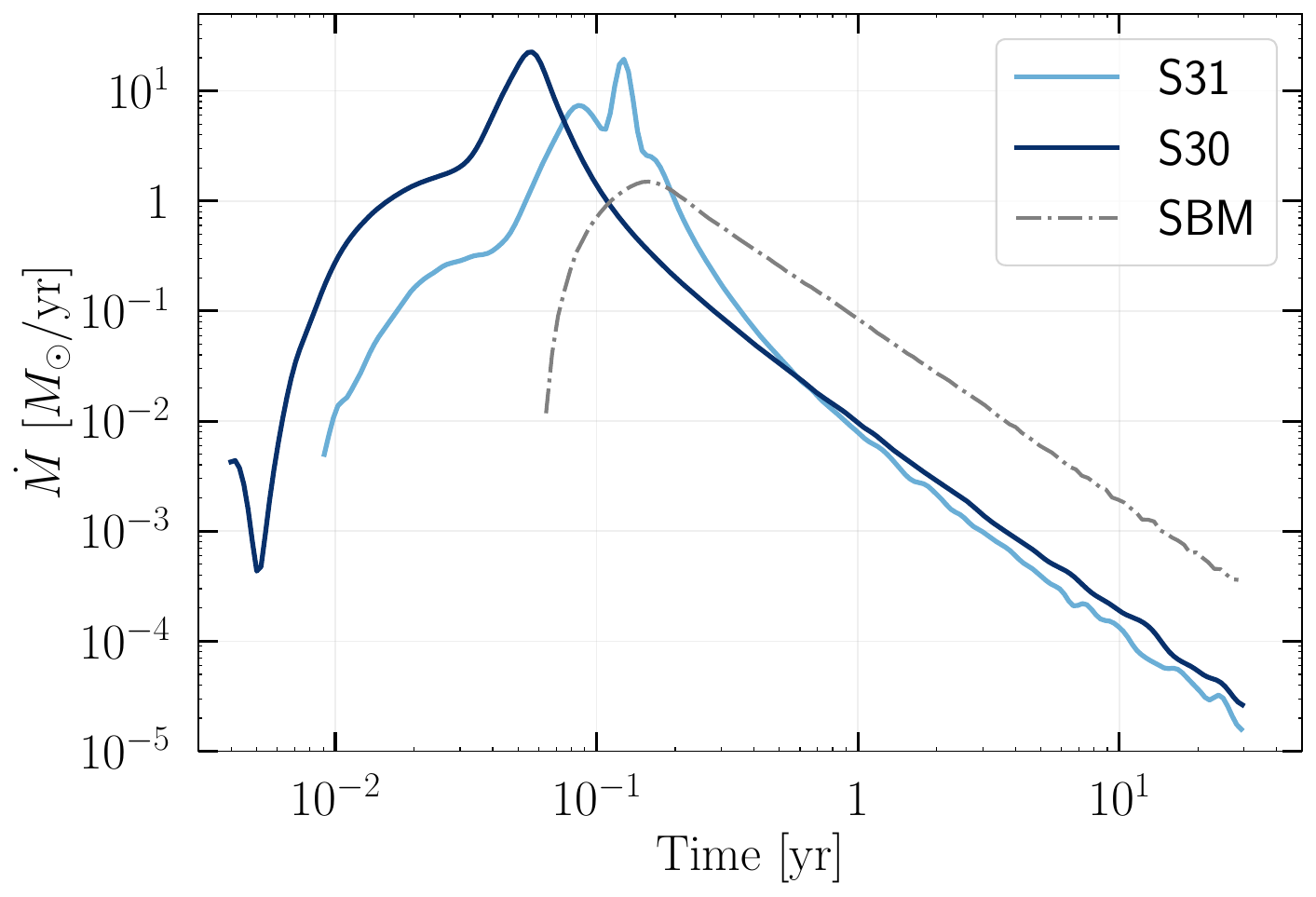}
    \caption{Fallback rate curves for the cTDEs with surviving SLS cores, shown for the representative cases S30 and S31. The gray dashed curve indicates the fallback rate of the single-star parabolic disruption (SBM) for reference. Both cases retain the early-time, broadened fallback morphology characteristic of collision-driven disruptions.}
    \label{fig:surviving_core}
\end{figure}

A small subset of cTDEs exhibit particularly unusual outcomes associated with the survival of a compact SLS core after the encounter. These events represent the rarest collision-driven configurations, accounting for less than one per cent of the total disruption population, and are confined to narrow regions of the collisional window. In particular, surviving bound cores are found for $\varphi_0\in(30^\circ,31^\circ)$, while surviving unbound cores appear for $\varphi_0\in(210^\circ,211^\circ)$. As discussed in Section~\ref{subsec:morphological_diversity}, these encounters produce a surviving stellar core that appears as a distinct isolated peak in the debris eccentricity distribution (e.g. middle panel of Figure~\ref{fig:reflejado}). In the bound cases, these peaks are concentrated around eccentricities of $e\sim0.98$, indicating that the remnant survives on a tightly bound orbit around the SMBH. The presence of this surviving core is also reflected in the fallback rate curves (Figure~\ref{fig:surviving_core}), where it produces additional narrow structures superimposed on the broader collision-driven fallback rate profile.

The fallback rate evolution of these events combines the signatures of strongly collision-driven disruptions with additional features associated with the surviving core. In particular, the fallback rate curves retain the very early onset and broadened morphology characteristic of cTDEs, while also developing narrow secondary structures at later times (e.g. S31), associated with the return of material bound to the surviving core. Such behaviour resembles repeating partial disruption scenarios, where a bound remnant undergoes multiple passages and episodic stripping events \citep{broggi2024repeating,chen2024fate,somalwar2025first}. Representative fallback rate curves for S30 and S31 are shown in the Figure~\ref{fig:surviving_core}.

Not all surviving cores remain bound to the SMBH. In some configurations, the surviving SLS core is instead ejected on an unbound trajectory together with a fraction of the debris (e.g. S211). In these cases, the fallback rate curve does not exhibit a clear secondary feature that can be unambiguously associated with the surviving core. Because similar behaviours in the fallback rate may also arise from variations in stellar or SMBH mass, these events may be difficult to uniquely identify observationally.

From a numerical standpoint, these cases are also among the most computationally demanding simulations in our sample. The surviving core has not fully relaxed by the end of the simulation runtime, preventing a robust determination of its final mass or radius. Instead, the core is identified as a clear overdensity embedded within the debris distribution.
\subsection{Event rates}
\label{subsec:rates}

The encounters studied here are a subset of binary tidal separations, so their
rate can be estimated by anchoring on the rate of the parent process rather than by
solving the loss-cone problem afresh. The separation rate is the same quantity that
sets the hypervelocity star production rate, for which \citet{yu2003ejection} obtain
$\Gamma_{\rm sep} \sim 10^{-5}\,(\eta/0.1)\ \mathrm{yr^{-1}}$ per galaxy, with $\eta$
the fraction of stars in binaries tight enough to be relevant ($a_0 \lesssim 0.3$ AU).
Our binary, with $a_0 = 11\,R_\odot \simeq 0.05$ AU, lies well inside this range, and
$\eta$ already absorbs the reduced binary fraction of a nuclear star cluster noted in
Section~\ref{sec:intro}. Notably, $\Gamma_{\rm sep}$ is comparable to the TDE rate itself.

Not every separation yields a disruption. The captured star reaches its own tidal
radius on the first passage only if $\beta_s = r_t/r_p > 1$, which in terms of the binary
impact parameter requires
\begin{equation}
\label{eq:beta_crit}
\beta_b > \beta_{b,{\rm crit}} = \frac{a_0}{r_1}\left(\frac{m_1}{m_b}\right)^{1/3} \simeq 8.7
\end{equation}
for our parameters, consistent with the value $\beta_s = 1.2$ obtained at $\beta_b = 10.5$.
Since the cross section for a gravitationally focused encounter scales linearly with
pericentre distance, the fraction of separations that are this deep is
$\sim \beta_{b,{\rm crit}}^{-1} \approx 0.1$. This is the full loss-cone scaling and should be taken as an upper limit, as deep penetrations are rarer in the diffusive regime \citep[for a review of loss-cone regimes see][]{stone2020rates}; strong scatterings
can further reduce the fraction of stars reaching the loss cone in the pinhole regime,
where deep penetrations are possible \citep{teboul2024}.

Combining the two factors gives $\Gamma \sim 10^{-6}\ \mathrm{yr^{-1}}$ per galaxy,
or roughly one per cent of the total TDE rate of $10^{-5}$--$10^{-4}\ \mathrm{yr^{-1}}$
per galaxy \citep{stone2016rates,vanvelzen2018}. Applying the phase fractions of
Table~\ref{tab:outcome_summary}, this divides into $\sim 5\times10^{-7}\ \mathrm{yr^{-1}}$
per galaxy for each of eTDEs and hTDEs, and $\sim 4\times10^{-8}\ \mathrm{yr^{-1}}$ per
galaxy for cTDEs. These figures are order-of-magnitude estimates: they inherit the
order-of-magnitude uncertainty in $\Gamma_{\rm sep}$ itself, and they assume that our
fiducial binary is representative.

We stress that this channel is not expected to dominate the eccentric TDE population.
As noted in Section~\ref{sec:ecc_TDE}, tidal capture operates at pericentre distances a
few times larger than $r_t$ and therefore has a considerably larger cross section, so it
is likely the more common route to a bound star. The mechanism considered here is instead
distinguished by being tractable hydrodynamically from first passage onwards, and by
producing signatures that tidal capture does not, in particular the WDDE and the
collision-driven events.

\section{Summary}
\label{sec:summary}

This section synthesizes the main findings of our study on the tidal disruption of stellar binary systems by a SMBH, highlighting how this channel naturally gives rise to a wide variety of exotic transient phenomena.

\subsection{Binary separation as a driver of eccentric TDEs}

While the tidal separation of stellar binaries is well known in stellar dynamics, its role in delivering TDE progenitors onto eccentric orbits has not been systematically explored hydrodynamically until this work. Our results demonstrate the robust efficiency of this pathway: the SLS is disrupted in every case, while the energy exchanged during the separation leaves its debris bound or unbound with equal probability, the WD/WDDE being correspondingly ejected as a hypervelocity object or retained on a tightly bound orbit.

Our simulations confirm that the resulting orbital eccentricities systematically deviate from the classical parabolic case ($e = 1$), leading to significant modifications in the fallback rates and temporal evolution of the debris. This establishes binary tidal separation as a physically motivated pathway for producing eccentric TDEs.

\subsection{Stellar collisions}

While three-body simulations capture the global energy exchange of the system, our results show that hydrodynamical effects are essential for understanding the most extreme outcomes. In particular, we identify two collisional windows in the initial phase $\varphi_0$ ($26^\circ \lesssim \varphi_0 \lesssim 32.5^\circ$ and $206^\circ \lesssim \varphi_0 \lesssim 212.5^\circ$) in which the WD directly impacts the SLS.
These collisions:
\begin{itemize}
    \item Broaden the eccentricity distribution of the debris, producing significantly earlier fallback rate times (as early as
 $\sim 0.034$ yr after pericenter),
    \item Enable the formation of a composite object consisting of the WD core with debris envelope (WDDE), capturing up to 60\% of the SLS mass in the most extreme cases;
more modest capture extends to the adjacent phases, $\varphi_0 \simeq 17^\circ$--$39^\circ$
and its $180^\circ$ complement, and is absent elsewhere,
    \item Eliminate the survival islands predicted by pure three-body simulations, as tidal forces and hydrodynamical interactions disrupt the binary.
\end{itemize}

\subsection{Classification of transients and observational signatures}

Our results reveal that binary--SMBH encounters can produce a diverse population of TDEs, including elliptical, hyperbolic, and partial disruption events. A particularly important outcome is that
hTDE configurations within the capture range are accompanied by a bound WDDE remnant, which may undergo repeated pericenter passages around the SMBH. In these systems, an initial hTDE powered by the returning debris can therefore be followed by subsequent accretion episodes associated with the WDDE, naturally producing delayed or repeating flaring activity.

The possible configurations are summarised schematically in Table~\ref{tab:encounter_schematic}, whose columns give the fate of the WD/WDDE and whose rows give the trajectories and fallback morphologies. Whether the WD/WDDE remains bound to the SMBH or is ejected as a hypervelocity object is the single property that most directly determines the observational appearance of the event, since it sets both the sign of the energy kick imparted to the SLS debris and whether recurrent activity is possible at later times.

This scenario offers two routes to recurrent activity. The broader one requires no envelope: wherever the WD/WDDE remains bound, which is half of all orientations, it continues to orbit through the debris and may perturb the accretion disc that forms at later times, providing a potential pathway to quasi-periodic eruptions (QPEs). The second is confined to the capture range, where a bound WDDE or a surviving SLS core retains material that can be partially stripped at each subsequent pericenter passage, producing repeating TDEs (rTDEs).

\begin{table*}
  \centering
  \caption{Schematic summary of the outcomes of a deep binary--SMBH encounter. The two
  columns give the fate of the WD/WDDE, which each account for exactly half of all binary
  orientations; the upper row shows the corresponding trajectories, the binary arriving in
  each case on the upper branch. The lower row gives the morphology of the fallback rate,
  an independent classification: the percentages within it refer to the fallback classes,
  which do not coincide with the WD/WDDE split, and are colour-matched to the curve they
  label. In both panels the black curve is the classic parabolic TDE, shown for reference.}
  \label{tab:encounter_schematic}
  \renewcommand{\arraystretch}{1.5}
  \begin{tabular}{|>{\raggedright\arraybackslash}m{0.12\textwidth}|
                  >{\centering\arraybackslash}m{0.36\textwidth}|
                  >{\centering\arraybackslash}m{0.36\textwidth}|}
    \hline
    &
    \textbf{Unbound WD/WDDE} ($50\%$) &
    \textbf{Bound WD/WDDE} ($50\%$) \\
    \hline
    \textbf{Trajectory} &
    \includegraphics{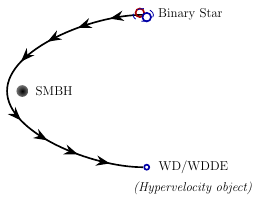} &
    \includegraphics{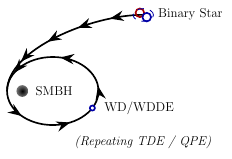} \\
    \hline
    \textbf{Fallback rate} &
    \includegraphics[width=\linewidth]{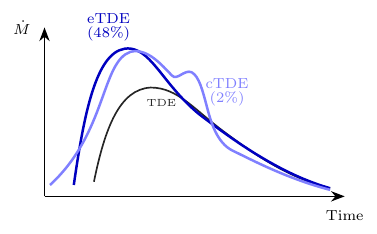} &
    \includegraphics[width=\linewidth]{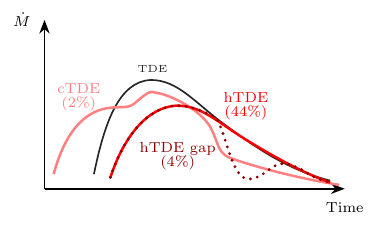} \\
    \hline
  \end{tabular}
\end{table*}

\subsection{Long-term evolution and final outcomes}

The long-term evolution of the WDDE opens several interesting astrophysical possibilities.
 In our simulations the WD accretes material only within the narrow capture windows, $\sim 12\%$ of all binary orientations, and for the innermost $\sim 1.5\%$ this brings the total mass of the WDDE above $1.4\,M_\odot$ while leaving the degenerate core below it.
However, given the rapid accretion timescales (of order hours), the system is unlikely to evolve to a Type Ia supernova. Instead, the most likely initial outcome is the occurrence of nova-like events driven by rapid mass transfer and envelope expansion, expected once the accretion rate onto the core exceeds the stable-burning rate of order $10^{-6}\,M_\odot\,\mathrm{yr^{-1}}$ (Section~\ref{subsec:wdde_fate}). Depending on how much mass is subsequently retained or expelled, the system may later evolve into a long-lived configuration resembling a peculiar red giant or, should enough of the envelope be burned onto the core for it to approach the Chandrasekhar mass, a delayed thermonuclear explosion.

\subsection{Model limitations}

We note that our study assumes coplanar encounters and treats the WD as a point mass due to its compact size. While this approximation allows us to resolve the hydrodynamics of the disrupted SLS with high accuracy, it neglects internal processes that might be relevant to the fate of the WDDE, such as nuclear reactions and thermal energy dissipation.

Furthermore, more complex scenarios involving non-coplanar geometries or binaries composed of two extended stars may lead to richer dynamical outcomes \citep[e.g.,][]{yu2025binary}. These limitations motivate future work exploring a broader parameter space and incorporating additional physical processes.

A further simplification is our neglect of the BH spin, which our Newtonian potential omits along with the spin dependence of the innermost stable circular orbit (ISCO). Lense-Thirring precession torques the returning debris out of the original orbital plane, which can delay stream self-intersection and lengthen the circularization timescale, in some cases by years \citep{guillochon2015}; this may matter more here than in the parabolic case, since the tightly bound debris produced by binary separation returns to pericenter far sooner. The ISCO radius also decreases from $6\,GM_{\rm BH}/c^2$ to $GM_{\rm BH}/c^2$ between a non-rotating and a maximally rotating BH in the prograde case, raising the late-time radiative efficiency by close to an order of magnitude and rescaling the luminosities quoted in Section~\ref{sec:menagerie}. Neither effect alters the debris energy distribution imprinted at pericentre, which is our primary result, but both bear on the observational appearance of these events.

\section{Conclusions}
\label{sec:conclusions}

We have shown that the tidal separation of stellar binaries, specifically those composed of a white dwarf (WD) and a solar-like star (SLS), by a SMBH provides a robust and natural mechanism for producing eccentric TDEs and a diverse array of exotic transients. This variant of the Hills mechanism, restricted to encounters deep enough that the captured star reaches its own tidal radius on the first passage, places one star on a tightly bound orbit while ejecting its companion, leading to systematic deviations from the canonical parabolic case and significantly altering the energetic budget, debris distribution and fallback rates.

Hydrodynamical effects play a critical role in shaping the most extreme outcomes. In particular, we identify two narrow collisional windows, together $\sim 4\%$ of all binary orientations, in orbital phase where direct WD--SLS collisions occur, triggering very early fallback rate and broadening the debris eccentricity distribution. These and the immediately adjacent phases, $\sim 12\%$ of orientations in total, encounters can also lead to the formation of a composite object consisting of a white dwarf surrounded by a debris envelope (WDDE); for the remaining $\sim 88\%$ the WD emerges unchanged.

These encounters naturally give rise to a range of transient classes, including elliptical TDEs (eTDEs), hyperbolic TDEs (hTDEs), collision-driven TDEs (cTDEs) and partial TDEs. They also open two routes to recurrent activity. The broader one requires no envelope: the WD/WDDE remains bound for half of all orientations, and may perturb the accretion disc that forms at later times, offering a pathway to quasi-periodic eruptions (QPEs). The second is confined to the capture range, where a bound WDDE or a surviving SLS core retains material that can be partially stripped at each subsequent pericenter passage, producing repeating TDEs (rTDEs).

In several cases, the WDDE accretes enough material
for its total mass to exceed $1.4\,M_\odot$, though the degenerate core remains below
that limit, pointing to a variety of possible outcomes, including nova-like episodes and
peculiar red giant-like objects.

Additionally, our results carry interesting implications for hypervelocity objects. When the WD/WDDE is not captured but instead ejected, it escapes at velocities of thousands of kilometers per second. Ejections from within the capture range carry an envelope of up to $\sim 0.6\,M_\odot$, whereas those from the remaining phases are bare WDs of unchanged mass. It is the presence of the envelope, rather than the mass itself, that offers a distinctive signature: a degenerate core wrapped in extended, cooling material would resemble neither an ordinary hypervelocity star nor a bare hypervelocity WD.

Finally, our results highlight binary--SMBH interactions as a fertile channel for generating exotic TDEs with observable signatures, motivating future studies that incorporate more general geometries and additional physical processes.

\section*{Acknowledgements}

We thank an anonymous referee for insightful comments that helped improve the clarity and presentation of this paper. We also thank Jane Arthur for valuable discussions. MG and SL acknowledge support from PAPIIT-UNAM grant IN102724. The authors gratefully acknowledge the computer resources, technical expertise and support provided by the Laboratorio Nacional de Supercómputo del Sureste de México (LNS), a member of the SECIHTI network of national laboratories. We acknowledge the use of Claude (Anthropic) for editorial assistance in improving the clarity and English grammar of specific sections of this manuscript. We reviewed and edited the output and take full responsibility for the final content.

\section*{Data Availability}
The data underlying this article will be shared on reasonable request to the corresponding author.


\bibliographystyle{mnras}
\bibliography{references}


\appendix

\section{Convergence Tests}
\label{app:convergence}

All convergence tests reported here correspond to the SPH simulation with initial binary phase $\varphi_0 = 32.5^\circ$ and impact parameter $\beta = 1.2$. Quantities are measured 53 hours after the pericenter passage unless otherwise stated.

\subsection{Resolution convergence}
\label{app:res_conv}

We performed a resolution study varying the number of SPH particles from 25,000 to 1,000,000 to verify that our main results do not depend on particle number. Table~\ref{tab:res_conv} summarizes the measured global quantities at $t = 53\,$hours after pericenter.

\begin{table*}
\centering
\caption{Resolution convergence test: global quantities measured 53 hours after pericenter.}
\label{tab:res_conv}
\begin{tabular}{cccccc}
\hline
Particles & Eccentricity (WD) & \% Bound particles & \% Unbound particles & Mass bound to the WD [$M_\odot$] \\
\hline
25,000 & 1.009372 & 90.956\% & 9.044\% & 1.062643 \\
50,000 & 1.007492 & 88.024\% & 11.976\% & 1.081024 \\
100,000 & 1.007932 & 89.378\% & 10.622\% & 1.073754 \\
500,000 & 1.007485 & 88.1946\% & 11.8054\% & 1.083402 \\
1000,000 & 1.007356 & 87.8824\% & 12.1176\% & 1.085763 \\
\hline
\end{tabular}
\end{table*}

\begin{figure}
    \centering
    \includegraphics[width=0.9\linewidth]{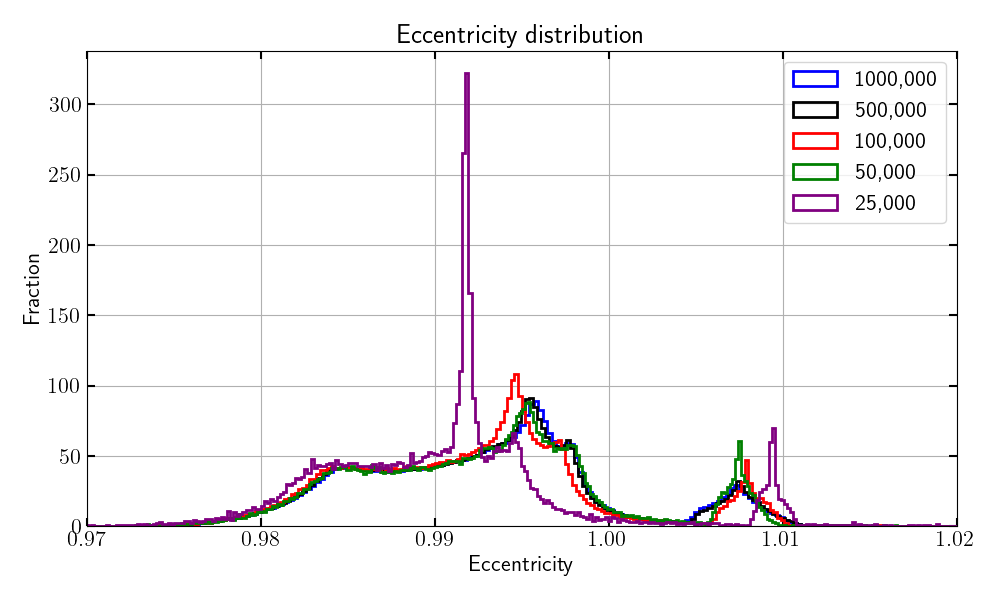}
    \caption{Final eccentricity distribution of the SLS for different resolutions: blue = 1,000,000 SPH, black = 500,000 SPH, red = 100,000 SPH, green = 50,000 SPH, purple = 25,000 SPH. All distributions correspond to $t=53\,$hours after pericenter.}
    \label{fig:A1}
\end{figure}

Figure~\ref{fig:A1} displays the final eccentricity histograms of the SLS for the different resolutions. The three highest-resolution runs (100k, 500k and 1M) produce nearly identical final eccentricity distributions, while the 50k run shows a small departure and the 25k run presents noticeable deviations. These results indicate that our standard choice of $N=500{,}000$ particles provides a converged description of the bulk dynamics on the timescale considered; the lowest-resolution case ($25{,}000$ particles) does not capture some details and deviates from the converged behaviour.

\subsection{Initial distance convergence}
\label{app:dist_conv}

We also tested the sensitivity of the outcome to the initial distance of the binary CM from the BH. All runs in this test used $N=100{,}000$ particles; initial distances were set to 10, 5 and 1 tidal radii ($R_t$). Table~\ref{tab:dist_conv} reports the measured quantities at $t = 53\,$hours after pericenter.

\begin{table*}
\centering
\caption{Convergence test with respect to the initial binary--BH distance (measured in tidal radius $R_t$). Quantities correspond to $t=53\,$hours after pericenter and were obtained with $N=100{,}000$ particles.}
\label{tab:dist_conv}
\begin{tabular}{cccccc}
\hline
Initial distance [$R_t$] & Eccentricity (WD) & Particles & \% Bound particles & \% Unbound particles & Mass bound to the WD [$M_\odot$]\\
\hline
10 & 1.00785 & 100,000 & 88.752\% & 11.248\% & 1.078584 \\
5 & 1.00779 & 100,000 & 88.997\% & 11.003\% & 1.076654 \\
1 & 1.00752 & 100,000 & 88.322\% & 11.678\% & 1.081064 \\
\hline
\end{tabular}
\end{table*}

\begin{figure}
    \centering
    \includegraphics[width=0.9\linewidth]{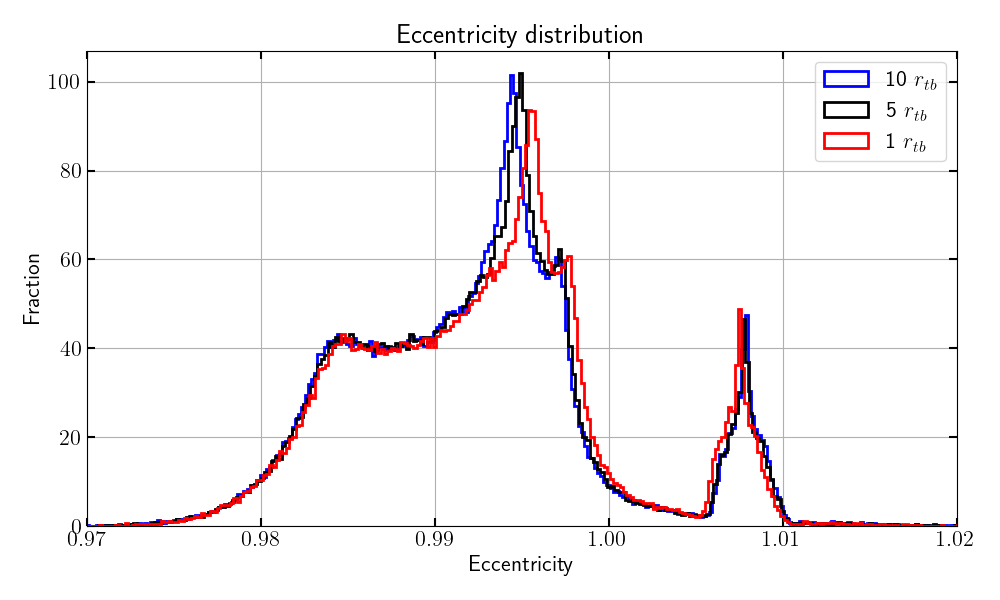}
    \caption{Final eccentricity distribution of the SLS for different initial distances: blue $=10\,R_t$, black = $5\,R_t$, red = $1\,R_t$. Distributions correspond to $t=53\,$hours after pericenter.}
    \label{fig:A2}
\end{figure}

Figure ~\ref{fig:A2} shows the final eccentricity distributions for the three initial distances. The distributions are virtually indistinguishable at t=53 hours after the pericenter: the eccentricity, energy fractions, and number of particles near the WD vary only slightly around the center of the distribution. This demonstrates that, within the range tested (1--10 $R_t$), the chosen starting distance does not significantly affect the dynamical evolution on the analyzed timescale.

\subsection{Simulation end time}
\label{app:time_conv}

To determine the duration required for the simulations, we monitored, for
the $\varphi_0 = 32.5^\circ$, $N = 500{,}000$ run, the fraction of stellar
debris gravitationally bound to the SMBH as a function of time after
pericenter, with particles classified according to the sign of their
specific orbital energy. The bound fraction drops rapidly at early times,
from $92.46\%$ at $t = 2.2\,$hours to $88.87\%$ at $t = 8.8\,$hours and
$88.25\%$ at $t = 24.2\,$hours, but changes by only $\simeq0.02$
percentage points between $t = 37.4\,$hours and $t = 72.6\,$hours,
indicating that the bound/unbound partition is essentially established
well before the end of the simulation. We therefore adopted a final
evolution time long enough to reach this quasi-steady state, without
following the subsequent long-term fallback evolution.
\subsection{Convergence summary}

Across the performed tests we conclude that:
\begin{itemize}
    \item The simulations are numerically converged for particle numbers $\gtrsim 100{,}000$ on the 53-hour timescale after pericenter. Differences between 100k, 500k and 1M are negligible for the diagnostics studied (eccentricity distribution, energy partition and particles near the WD).
    \item The lowest-resolution run (25k) shows non-negligible deviations and should be considered under-resolved for the phenomena reported in this work.
    \item The choice of initial starting distance in the range $1$--$10\,R_t$ does not significantly affect the main outcomes at $t=53\,$hours, validating our default choice of initial separation used in the production runs.
\end{itemize}
\section{Calculation of the mass bound to the WD}
\label{app:bound_mass_wd}

In this appendix we describe the procedure used to estimate the mass of gas gravitationally bound to the WD ($m_2$) in our hydrodynamical simulations of binary--SMBH encounters. The analysis is performed using the final time of each simulation, corresponding to approximately $53\,\mathrm{h}$ after pericentre passage.

\subsection{Identification of bound gas}

For each snapshot, we first identify the point-particle associated with the WD, whose mass in our simulations is
\begin{equation}
m_2 = 1\,M_\odot.
\end{equation}

The positions and velocities of all gas particles are then expressed in the reference frame of the WD by computing the relative quantities
\begin{equation}
\mathbf{r}'_i = \mathbf{r}_{{\rm gas},i} - \mathbf{r}_2, \qquad
\mathbf{v}'_i = \mathbf{v}_{{\rm gas},i} - \mathbf{v}_2,
\end{equation}
where $(\mathbf{r}_2, \mathbf{v}_2)$ are the position and velocity of the WD.

To determine whether a particle is gravitationally bound to the WD we compute its specific orbital energy
\begin{equation}
E_i = \frac{1}{2}v_i'^2 - \frac{G m_2}{r'_i},
\end{equation}
and its specific angular momentum
\begin{equation}
\mathbf{h}_i = \mathbf{r}'_i \times \mathbf{v}'_i, \qquad h_i = |\mathbf{h}_i|.
\end{equation}

From these quantities we obtain the orbital eccentricity,
\begin{equation}
e_i = \sqrt{1 + \frac{2 E_i h_i^2}{(G m_2)^2}}.
\end{equation}

A gas particle is considered gravitationally bound to the WD if it satisfies the condition $e_i < 1$, corresponding to a bound (elliptical) orbit around the WD. Particles with non-physical values caused by numerical noise are excluded from the analysis.

The total mass dynamically bound to the WD is then obtained by summing the masses of all particles fulfilling this criterion,
\begin{equation}
M_{\rm bound,2} = \sum_{i,\, e_i < 1} m_i,
\end{equation}
where $m_i$ is the mass of each gas particle. This quantity represents the total amount of stellar material captured by the WD at the end of the simulation.

\subsection{Characteristic radius of the WDDE}
\label{app:wdde_radius}

We define the White Dwarf with Debris Envelope (WDDE) as the composite object formed by the WD ($m_2$) and the gas gravitationally bound to it. To characterise its spatial extent, we define a characteristic radius that encloses a fixed fraction of the bound mass.

Using the same set of bound particles (i.e., those satisfying $e_i < 1$), we consider their radial distances from the WD in its rest frame,
\begin{equation}
r'_i = |\mathbf{r}'_i|.
\end{equation}

We then construct the cumulative mass distribution by sorting the particles according to $r'_i$. Since all gas particles have equal mass in our simulations, this is equivalent to ordering them by radius and counting particles.

The characteristic radius of the WDDE is defined as the radius enclosing $95\%$ of the total bound mass,
\begin{equation}
R_{95} \equiv r'_{(k)}, \qquad k = \lfloor 0.95\,N_{\rm bound} \rfloor,
\end{equation}
where $N_{\rm bound}$ is the total number of bound particles and $r'_{(k)}$ denotes the $k$-th element of the sorted radial distances.

By construction, the enclosed mass within $R_{95}$ satisfies
\begin{equation}
M(<R_{95}) = 0.95\,M_{\rm bound,2}.
\end{equation}

This definition provides a robust, resolution-independent estimate of the WDDE size that is insensitive to low-density material at large radii. In all cases, $R_{95}$ is computed at the final simulation time, consistently with the bound-mass estimate.
\section{Comparison with three-body dynamics}
\label{app:comparison_3body}

Our hydrodynamical simulations show both a remarkable agreement and critical deviations with respect to the predictions of three-body dynamics.

A fundamental difference arises from the hydrodynamical treatment: in sufficiently close encounters, the WD can physically collide with the SLS or with its expanding debris. Such interactions are inherently absent in the purely gravitational three-body framework, where both stars are modeled as point masses. As identified in Section~\ref{subsec:three_body}, within the critical range of initial orbital phase $26^\circ \lesssim \varphi_0 \lesssim 32.5^\circ$ and $206^\circ \lesssim \varphi_0 \lesssim 212.5^\circ$, direct collisions between the WD and the SLS occur at pericenter. Outside this interval, the WD can still interact with the extended debris stream at later times, once the stellar material has undergone significant tidal expansion after disruption.

The level of agreement with the three-body predictions is directly determined by the collision geometry. For encounters outside the critical phase ranges, the trajectories remain in excellent agreement for most of the orbital evolution, with only minor deviations appearing after the pericenter passage.

\begin{figure}
    \centering
    \includegraphics[width=1.0\linewidth]{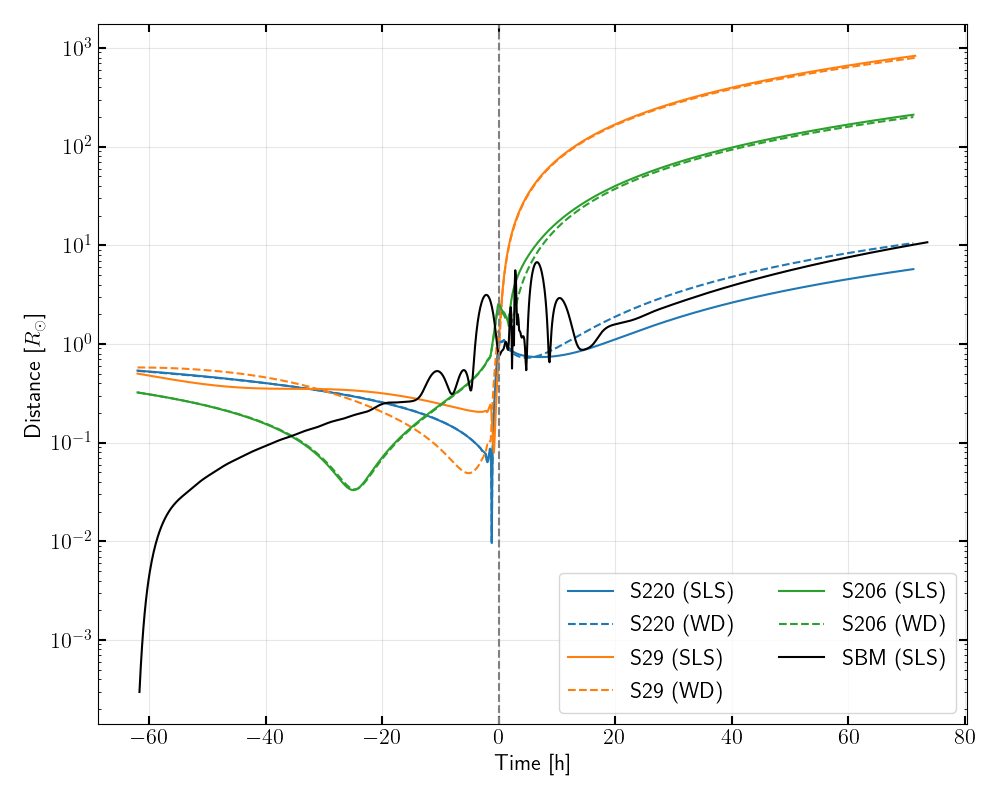}
    \caption{Time evolution of the relative separation between the trajectories obtained from the hydrodynamical simulations and from the three-body model. Dashed lines correspond to the WD, while solid lines show the center-of-mass motion of the SLS. The black curve represents the single-star reference case (SBM).}
    \label{fig:trajectory_separation}
\end{figure}

Figure~\ref{fig:trajectory_separation} shows the time evolution of the trajectory separation for representative simulations, defined as
\begin{equation}
D(t) = \left|\mathbf{r}_{\text{hydro}}(t) - \mathbf{r}_{\text{3body}}(t)\right|,
\end{equation}
where $\mathbf{r}_{\text{hydro}}$ and $\mathbf{r}_{\text{3body}}$ correspond to the WD position in the hydrodynamical and three-body simulations, respectively. The dashed curves show the relative separation between the WD trajectories in both approaches, while the solid curves represent the separation between the center-of-mass trajectory of the SLS in the hydrodynamical simulation and the corresponding point-mass trajectory in the three-body model. The black curve shows the reference single-star case (SBM).

Before the pericenter passage, both descriptions follow essentially the same trajectory, with differences smaller than one solar radius. However, after pericenter the trajectories diverge significantly. The primary cause of this deviation in the hydrodynamical simulations is the tidal disruption of the SLS. Nevertheless, the magnitude of the departure from the three-body solution also depends on the strength of the stellar interaction at pericenter.

The largest deviation is observed for simulation S29, which corresponds to a nearly head-on collision. This is followed by case S206, where a collision still occurs at pericenter but with a less direct geometry. In contrast, in simulation S220 the stars do not collide at pericenter and the separation between the hydrodynamical and three-body trajectories remains below $\sim 10\,R_\odot$, indicating only a modest deviation. The black curve corresponding to the single-star case lies within the same range.

These results are consistent with the interpretation that the more frontal the collision at pericenter, the larger the deviation from the trajectory predicted by three-body dynamics.


\bsp
\label{lastpage}
\end{document}